\documentclass{article}

\usepackage{amsmath}
\usepackage{amsfonts}
\usepackage{amssymb}
\usepackage{mathtools}
\usepackage{amsthm}
\usepackage{bm}
\usepackage{hyperref}
\usepackage{titlesec}
\usepackage{array}
\usepackage{epstopdf}
\usepackage{authblk}
\usepackage{fullpage}
\usepackage{setspace}
\usepackage{algorithm}
\usepackage{algpseudocode}
\usepackage{url}
\usepackage{caption, subcaption}
\usepackage{graphicx}
\usepackage{fixmath}
\usepackage{color}
\usepackage{stackrel}
\usepackage{booktabs}

\newcommand{\tensor}[1]{\boldsymbol{#1}}
\newcommand{\mexp}[1]{\mathrm{E}\left\{#1\right\}}
\newcommand{\trans}{\top}
\newcommand{\cov}{\text{Cov}}

\newcommand{\dif}{\mathrm{d}}
\newcommand{\me}{\mathrm{e}}

\numberwithin{equation}{section}

\makeatletter
\def\mathcenterto#1#2{\mathclap{\phantom{#1}\mathclap{#2}}\phantom{#1}}
\let\old@widetilde\widetilde
\def\widetildeto#1#2{\mathcenterto{#2}{\old@widetilde{\mathcenterto{#1}{#2\,}}}}
\makeatother

\newcommand{\wtl}{\widetildeto{I}{L}}

\newcommand{\wty}{\widetildeto{y}{\bm y}}
\newcommand{\wtm}{\widetildeto{y}{\bm\mu}}
\newcommand{\wtc}{\widetildeto{y}{\bm c}}
\newcommand{\wtC}{\widetildeto{C}{\tensor C}}

\newcommand{\PreserveBackslash}[1]{\let\temp=\\#1\let\\=\temp}
\newcolumntype{C}[1]{>{\PreserveBackslash\centering}p{#1}}

\allowdisplaybreaks[1]

\title{Multifidelity Data Fusion via Gradient-Enhanced Gaussian Process Regression}
\author[1,2]{Yixiang Deng}
\author[3,4]{Guang Lin}
\author[5]{Xiu Yang\thanks{xiy518@lehigh.edu}}
\affil[1]{School of Engineering, Brown University, USA}
\affil[2]{Division of Applied Mathematics, Brown University, USA}
\affil[3]{Department of Mathematics, Purdue University, USA} 
\affil[4]{School of Engineering, Purdue University, USA} 
\affil[5]{Department of Industrial and Systems Engineering, Lehigh University, USA}

\begin{document}

\maketitle


\begin{abstract}

We propose a data fusion method based on multi-fidelity Gaussian process regression (GPR) framework. This method combines available data of the quantity of interest (QoI) and its gradients with different fidelity levels, namely, it is a Gradient-enhanced Cokriging method (GE-Cokriging). It provides the approximations of both the QoI and its gradients \emph{simultaneously} with uncertainty estimates.
We compare this method with the conventional multi-fidelity Cokriging method that does not use gradients information, and the result suggests that GE-Cokriging has a better performance in predicting both QoI and its gradients. Moreover, GE-Cokriging even shows better generalization result in some cases where Cokriging performs poorly due to the singularity of the covariance matrix. We demonstrate the application of GE-Cokriging in several practical cases including reconstructing the trajectories and velocity of an underdamped oscillator with respect to time simultaneously, and investigating the sensitivity of power factor of a load bus with respect to varying power inputs of a generator bus in a large scale power system. We also show that though GE-Cokriging method requires a little bit higher computational cost than Cokriging method, the result of accuracy comparison shows that this cost is usually worth it.
\end{abstract}

\maketitle


\section{Introduction}
\label{sec:intro}

Gaussian process (GP) is one of the most well studied stochastic processes in probability and statistics. Given 
the flexible form of data representation, GP is a powerful tool for classification and regression, and it is  
widely used in probabilistic scientific computing, engineering design, geostatistics, data assimilation, machine learning, etc. In particular, given a data set comprising input/output pairs of locations and quantity of interest (QoI),
\emph{GP regression} (GPR, also known as \emph{Kriging}), can provide a prediction along with a mean squared error (MSE) estimate of the QoI at any location. Alternatively, from the Bayesian perspective, GPR identifies a Gaussian random variable at any location with a posterior mean (corresponding to the prediction) and variance (corresponding to the MSE). Generally speaking, the larger the given data set size is, the closer the GPR's posterior mean is to the ground truth and the smaller the posterior variance is.

In many practical problems, obtaining a large amount of data can be difficult because of the limitation of resources. There are several approaches to augment the data set in different manners. For example, the original Cokriging method exploits the correlation between multiple QoIs in the geostatistical study, e.g., the correlation between temperature and precipitation~\cite{goovaerts1998ordinary, stein1991simulation, stein1991universal}, or that between near-surface soil density and the gravity-gradient~\cite{geng20143d}, to improve the accuracy of prediction. Later, the Cokriging method was extended to utilizing correlation between the same QoI from models with different fidelities~\cite{kennedy2000predicting, girolami2007data, le2014recursive, perdikaris2015multi}. This GP-based multi-fidelity method is very useful in scientific computing, because low-fidelity models, e.g., coarse-grained molecular dynamics~\cite{espanol1995statistical, rudd1998coarse}, Reynolds-average Navier-Stokes equations~\cite{reynolds1895iv, alfonsi2009reynolds}, numerical simulations on coarse grids, are often used with high-fidelity models, e.g., molecular dynamics, full Navier-Stokes equations, numerical simulations on fine grids~\cite{lee2017general}, in optimization, uncertainty quantification (UQ), control~\cite{peherstorfer2018survey}, variable-fidelity quantum mechanical calculations of bandgaps of solids~\cite{pilania2017multi}, etc. In these tasks, the multi-fidelity method leverages low-fidelity models for speedup, while uses a high-fidelity model to establish accuracy and/or convergence guarantees. Moreover, the empirical statistics of simulation results from stochastic scientific computing models can be used to construct single- or multi-fidelity GP models~\cite{yang2018physics, yang2019physics, yang2020bifidelity}.
In this work, Cokriging refers to the GP-based multi-fidelity approach.



Another important approach to enlarge the data set is to use gradient information of the QoI. This approach can be categorized as Cokriging because the QoI and its gradients are variables of different species.
The idea of incorporating derivatives or gradients to optimize Bayesian prediction was proposed by Morris et al.~\cite{morris1993bayesian}. 
The \emph{gradient-enhanced Kriging} (GE-Kriging) method, also referred to as Gradient-based Kriging in some literature, has been widely investigated in areas such as computational fluid dynamics, especially in aerodynamics optimization problems~\cite{dwight2009efficient, xuan2009gradient, laurenceau2010comparison, chung2002design}. Incorporating gradient information in different ways, this method consists of direct and indirect approaches. 
The former uses the gradient information through an augmented covariance matrix~\cite{han2013improving}, while 
the latter approximates the gradient via finite-difference method~\cite{chung2002design,zimmermann2013maximum}.
The \emph{gradient-enhanced Cokriging} (GE-Cokriging) method in~\cite{laurent2019overview} refers to a GE-Kriging method that uses a different covariance function between the QoI and its gradients other than that in conventional GE-Kriging. The GE-Cokriging method in~\cite{ulaganathan2015performance} combines multi-fidelity information of the QoI and its gradients to predict the QoI only.  

Most of the aforementioned works focus on enhancing the accuracy of predicting the QoI. Hence, when the gradient information is used, the method is a ``gradient-enhanced'' approach. However, in many applications, both the QoI and its gradient are important. For example, when studying the phase diagram of a dynamical system, one needs an accurate prediction of both location and velocity. Another example is the sensitivity analysis of a system, where the gradient information is critical. Therefore, in this work, we propose a comprehensive 
\emph{multifidelity gradient-enhanced Cokriging} method
to predict both QoI and its gradients \emph{simultaneously} based on GE-Cokriging~\cite{ulaganathan2015performance}.
This method exploits the QoI and its gradient from models of different fidelities based on the combination of the GE-Kriging and the Cokriging to improve the prediction accuracy.
In terms of predicting the QoI, this method can be considered as ``gradient-enhanced'', while from the perspective of estimating gradients, this method can be considered as ``integral-enhanced''. In this work, GE-Cokriging refers to our proposed multi-fidelity method, instead of the GE-Cokriging in~\cite{laurent2019overview}.

In this paper, we firstly review GPR (Kriging) and its extension for a multi-fidelity study (Cokriging). Then, we describe the gradient-enhanced Kriging/Cokriging as well as the GE-Kriging/Cokriging method. Finally, we use four examples to demonstrate the efficacy of our approach. 

\section{Methodology}
\label{sec:method}


\subsection{GPR framework}
\label{subsec:gpr}
We present a brief review of the GPR method adopted from~\cite{abrahamsen1997review, forrester2008engineering, yang2018physics}.
We denote the observation locations as $\bm X = \{\bm x^{(i)}\}_{i=1}^N$
($\bm x^{(i)}\in D, D\subseteq\mathbb{R}^d$) and the 
observed values of the QoI at these locations as 
$\bm y=(y^{(1)}, y^{(2)},\dotsc, y^{(N)})^\trans$
($y^{(i)}\in\mathbb{R}$). For simplicity, we assume that $y^{(i)}$ are scalars. 
The GPR method aims to identify a GP $Y(\bm x,\omega):D\times\Omega\rightarrow\mathbb{R}$ based on the input/output data set $\{(\bm x^{(i)}, y^{(i)})\}_{i=1}^N$, where $\Omega$ is the sample space of a probability triple.
Here, $\bm x$ can be considered as parameters for this GP, such that 
$Y(\bm x,\cdot):\Omega\rightarrow \mathbb{R}$ is a Gaussian random 
variable for any $\bm x$ in the set $D$. A GP $Y(\bm x, \omega)$ is usually denoted as
\begin{equation}
  \label{eq:gp0}
Y(\bm x) \sim \mathcal{GP}\left(\mu(\bm x), k(\bm x, \bm x')\right),
\end{equation}
where $\omega$ is not explicitly listed for brevity, $\mu(\cdot):D\rightarrow\mathbb{R}$ and 
$k(\cdot,\cdot):D\times D\rightarrow\mathbb{R}$ are the mean 
and covariance functions (also called \emph{kernel} function), respectively:
  \begin{align}
    \mu(\bm x) & = \mexp{Y(\bm x)},\\ 
    k(\bm x,\bm x') & = \cov\left\{Y(\bm x), Y(\bm x')\right\}
                      = \mexp{(Y(\bm x)-\mu(\bm x))(Y(\bm x')-\mu(\bm x'))}.
  \end{align}
The variance of $Y(\bm x)$ is $k(\bm x, \bm x)$, and its standard deviation is
$\sigma(\bm x)=\sqrt{k(\bm x,\bm x)}$. 
The covariance matrix, denoted as $\tensor C$, is 
defined as $C_{ij}=k(\bm x^{(i)}, \bm x^{(j)})$.
Functions $\mu(\bm x)$ and $k(\bm x,\bm x')$ are obtained by identifying their
hyperparameters via maximizing the log marginal 
likelihood~\cite{williams2006gaussian}:
\begin{equation}
  \label{eq:lml}
  \ln L=-\dfrac{1}{2}(\bm y-\bm\mu)^\trans \tensor C^{-1} (\bm
  y-\bm\mu)-\dfrac{1}{2}\ln |\tensor C|-\dfrac{N}{2}\ln 2\pi,
\end{equation}
where $\bm\mu=(\mu(\bm x^{(1)}),\dotsc,\mu(\bm x^{(N)}))^\trans$ and $\vert\tensor C\vert$ is the determinant of matrix $\tensor C$. For any $\bm
x^*\in D$, the GPR posterior mean and variance are
  \begin{align}
\label{eq:krig}
\hat y(\bm x^*) & = \mu(\bm x^*) + 
  \bm c(\bm x^*)^\trans\tensor{C}^{-1}(\bm y-\bm\mu),  \\
  \hat s^2(\bm x^*) & = 
   \sigma^2(\bm x^*)-\bm c(\bm x^*)^\trans \tensor{C}^{-1}\bm c(\bm x^*),
 \end{align}
where $\bm c(\bm x^*)$ is a vector of covariance:
$(\bm c(\bm x^*))_i=k(\bm x^{(i)},\bm x^*)$.
In practice, it is common to use $\hat y(\bm x^*)$ as the prediction, and
$\hat s^2(\bm x^*)$ is also called the mean squared error (MSE) of the prediction
because $\hat s^2(\bm x^*)=\mexp{(\hat y(\bm x^*)-Y(\bm x^*))^2}$
\cite{forrester2008engineering}. Consequently, $\hat s(\bm x^*)$, the posterior standard deviation, is called the root
mean squared error (RMSE). 
Moreover, to account for the observation noise, one can assume that the noise is
independent and identically distributed (i.i.d.) Gaussian random variables with
zero mean and variance $\delta^2$, and replace $\tensor C$ with 
$\tensor C+\delta^2\tensor I$. In this study, we assume that observations 
$\bm y$ are noiseless. If $\tensor C$ is not invertible or its condition number
is very large, one can add a small regularization term $\alpha\tensor I$
($\alpha$ is a small positive real number) to $\tensor C$, 
which is equivalent to assuming there
is an observation noise. In addition, $\hat s$ can be used in global
optimization, or in the greedy algorithm to identify locations of additional
observations. 


\subsection{Kriging and Cokriging with stationary kernel}
\label{subsec:stationary_gpr}
In the widely used ordinary Kriging method, a stationary GP is assumed
\cite{kitanidis1997introduction}. Specifically, $\mu$ is set as a constant 
$\mu(\bm x)\equiv\mu$, and $k(\bm x, \bm x')=k(\bm\tau)$, where 
$\bm\tau=\bm x-\bm x'$. Consequently, 
$\sigma^2(\bm x)=k(\bm x,\bm x)=k(\bm 0)=\sigma^2$ is a constant.
The most widely used kernels in scientific computing is the Mat\'{e}rn functions, especially its two special cases, i.e., exponential and
squared-exponential (Gaussian) kernels. For example, the Gaussian 
kernel can be written as
$k(\bm\tau)=\sigma^2\exp\left(-\frac{1}{2}\Vert \bm x-\bm x'\Vert^2_w\right)$,
where the weighted norm is defined as 
$\displaystyle\Vert \bm x-\bm x'\Vert^2_w=\sum_{i=1}^d \left(\dfrac{x_i-x'_i}{l_i}\right)^2$. 
Here, $l_i$ ($i=1,\dotsc, d$), the correlation lengths in
the $i$ direction, are constants. Given a stationary covariance function, the
covariance matrix $\tensor C$ can be written as
$\tensor C=\sigma^2\tensor \Psi$, where
$\Psi_{ij}=\exp(-\frac{1}{2}\Vert\bm x^{(i)}-\bm x^{(j)}\Vert_w^2)$. The estimators of $\mu$ and $\sigma^2$, denoted as $\hat\mu$ and
$\hat\sigma^2$, are
\begin{equation}
\label{eq:krig_est}
   \hat\mu=\dfrac{\bm 1^\trans \tensor\Psi^{-1}\bm y}
                  {\bm 1^\trans\tensor\Psi^{-1}\bm 1}, \qquad
   \hat\sigma^2=\dfrac{(\bm y-\bm 1\hat\mu)^\trans\tensor\Psi^{-1}(\bm y-\bm 1\hat\mu)}{N},
\end{equation}
where $\bm 1$ is a constant vector consisting of 
$1$s~\cite{forrester2008engineering}. It is also common to set $\mu=0$~\cite{williams2006gaussian}. The hyperparameters $\sigma$ and $l_i$ are 
identified by maximizing the log marginal likelihood in Eq.~\eqref{eq:lml}. The terms
$\hat y(\bm x^*)$ and $\hat s^2(\bm x^*)$ in Eq.~\eqref{eq:krig} take the 
following form:
\begin{align}
\label{eq:krig_mean}
\hat y(\bm x^*) & = \hat\mu + \bm\psi^\trans\tensor\Psi^{-1}(\bm y-\bm 1\hat\mu), \\
\label{eq:krig_var}
\hat s^2(\bm x^*) & = \hat\sigma^2\left(1-\bm\psi^\trans\tensor\Psi^{-1}\bm\psi\right),
\end{align}
where $\bm\psi=\bm\psi(\bm x^*)$ is a (column) vector consisting of correlations between the observed data and the
prediction, i.e., $\psi_i=\frac{1}{\sigma^2}k(\bm x^{(i)},\bm x^*)$.

Next, we briefly review the formulation of the multifidelity Cokriging, and we use the two-fidelity model for demonstration. Suppose that we have high-fidelity data 
(e.g., accurate measurements of the QoI) 
$\bm y_{H}=(y_{H}^{(1)}, \dotsc,y_{H}^{(N_H)})^\trans$ at 
locations $\bm X_{H} = \{\bm x_{H}^{(i)}\}_{i=1}^{N_H}$, and 
low-fidelity data (e.g., measurements with lower accuracy or numerical approximations of the QoI) 
$\bm y_{L}=(y_{L}^{(1)}, \dotsc,y_{L}^{(N_L)})^\trans$ at 
locations $\bm X_{L} = \{\bm x_{L}^{(i)}\}_{i=1}^{N_L}$, where 
$y_{H}^{(i)}, y_{L}^{(i)}\in\mathbb{R}$ and 
$\bm x_{H}^{(i)}, \bm x_{L}^{(i)}\in D\subseteq\mathbb{R}^d$.
We denote $\bm X=\{\bm X_{L}, \bm X_{H}\}$ and 
$\wty = (\bm y_{L}^{\trans}, \bm y_{H}^{\trans} )^{\trans}$. 
Kennedy and O'Hagan \cite{kennedy2000predicting} proposed a multifidelity 
formulation based on the auto-regressive model for GP $Y_{H}(\cdot)$
($\sim\mathcal{GP}(\mu_{H}(\cdot), k_{H}(\cdot,\cdot))$):
\begin{equation}
  \label{eq:arm}
  Y_{H}(\bm x) = \rho Y_{L}(\bm x) + Y_{d}(\bm x),
\end{equation}
where $Y_{L}(\cdot)$ ($\sim\mathcal{GP}(\mu_{L}(\cdot), k_{L}(\cdot, \cdot))$) 
regresses the low-fidelity data, $\rho \in \mathbb{R}$ is a
regression parameter and $Y_d(\cdot)$ 
($\sim\mathcal{GP}(\mu_{d}(\cdot), k_{d}(\cdot,\cdot))$) models the discrepancy 
between $Y_{H}$ and $\rho Y_{L}$. This model assumes that
\begin{equation}
  \label{eq:arm-cov}
  \cov \left\{Y_{H}(\bm x), Y_{L}({\bm x}') \mid Y_{L}(\bm x) \right\}=0,
  \quad \text{for all}\quad \bm x' \neq \bm x, \ \bm x, \bm x' \in D.
\end{equation}
The covariance of observations, $\wtC$, is then given by
\begin{equation}
  \label{eq:cokrig_cov}
  \wtC=
  \begin{pmatrix}
    \tensor C_{L}(\bm X_{L}, \bm X_{L}) & \rho \tensor C_{L}(\bm X_{L}, \bm X_{H}) \\
    \rho \tensor C_{L} (\bm X_{H}, \bm X_{L}) & \rho^2 \tensor C_{L}(\bm
    X_{H}, \bm X_{H}) + \tensor C_{d}(\bm X_{H}, \bm X_{H})
  \end{pmatrix},
\end{equation}
where $\tensor C_{L}$ and $\tensor C_{d}$ are the covariance matrices computed
from $k_{L}(\cdot,\cdot)$ and $k_{d}(\cdot,\cdot)$, respectively, i.e.,
\begin{equation}
\label{eq:cokrig_ele}
\begin{aligned}
& [\tensor C_{L}(\bm X_{L}, \bm X_{L})]_{ij}=k_L(\bm X_L^{(i)}, \bm X_L^{(j)}), & [\tensor C_{L}(\bm X_{L}, \bm X_{H})]_{ij}=k_L(\bm X_L^{(i)}, \bm X_H^{(j)}),\\
& [\tensor C_{L}(\bm X_{H}, \bm X_{L})]_{ij}=k_L(\bm X_H^{(i)}, \bm X_L^{(j)}), & [\tensor C_{L}(\bm X_{H}, \bm X_{H})]_{ij}=k_L(\bm X_H^{(i)}, \bm X_H^{(j)}), \\
& [\tensor C_{d}(\bm X_{H}, \bm X_{H})]_{ij}=k_d(\bm X_H^{(i)}, \bm X_H^{(j)}). &
\end{aligned}
\end{equation}

One can assume parameterized forms for these kernels (e.g., Gaussian kernel) and employ 
the following two-step approach~\cite{forrester2007multi,
forrester2008engineering} to identify hyperparameters:
\begin{enumerate}
\item Use Kriging to construct $Y_{L}$ based on $\{\bm X_{L}, \bm y_{L}\}$. 
\item Denote $\bm y_{d} = \bm y_{H} - \rho \bm y_{L}(\bm X_{H})$, where 
$\bm y_{L}(\bm X_{H})$ are the values of $\bm y_{L}$ at locations common to
those of $\bm X_{H}$, then construct $Y_d$ using $\{\bm X_{H}, \bm y_{d}\}$ 
via Kriging. 
\end{enumerate}
The posterior mean and variance of $Y_H$ at $\bm x^*\in D$ are given by
\begin{align}
  \label{eq:cokrig_mean}
  \hat y(\bm x^*) & =  \mu_{H}(\bm x^*) + 
      \wtc(\bm x^*)^\trans\wtC^{-1}(\wty-\wtm),  \\
  \label{eq:cokrig_var}
      \hat s^2(\bm x^*) & = \rho^2\sigma^2_{L}(\bm x^*) + \sigma^2_{d}(\bm x^*)
      - \wtc(\bm x^*)^\trans \wtC^{-1}\wtc(\bm x^*),
\end{align}
where $\mu_{H}(\bm x^*)=\rho\mu_{L}(\bm x^*)+\mu_d(\bm x^*)$, 
$\sigma^2_{L}(\bm x^*)=k_{L}(\bm x^*, \bm x^*)$,
$\sigma^2_d(\bm x^*)=k_d(\bm x^*, \bm x^*)$, and
\begin{align}
  \label{eq:cokrig_mu}
\wtm & = \begin{pmatrix}\bm\mu_{L}\\ \bm\mu_{H}\end{pmatrix}
  = \begin{pmatrix}\left(\mu_{L}(\bm x_{L}^{(1)}),\dotsc, 
    \mu_{L}(\bm x_{L}^{(N_{L})})\right)^\trans \\
     \left(\mu_{H}(\bm x_{H}^{(1)}),\dotsc, 
\mu_{H}(\bm x_{H}^{(N_{H})})\right)^\trans\end{pmatrix},  \\
  \label{eq:cokrig_c}
  \wtc(\bm x^*)& = \begin{pmatrix}\rho\bm c_{L}(\bm x^*)\\ \bm c_{H}(\bm
x^*)\end{pmatrix}
= \begin{pmatrix}\left(\rho k_{L}(\bm x_{L}^{(1)},\bm x^*),\dotsc,
  \rho k_{L}(\bm x_{L}^{(N_{L})},\bm x^*)\right)^\trans \\
  \left(k_{H}(\bm x_{H}^{(1)},\bm x^*),\dotsc,k_{H}(\bm x_{H}^{(N_{H})},\bm x^*)
\right)^\trans\end{pmatrix}, 
\end{align}
where
$k_{H}(\bm x,\bm x') = \rho^2k_{L}(\bm x,\bm x') + k_{d}(\bm x, \bm x')$.
Alternatively, one can simultaneously identify hyperparameters in
$k_{L}(\cdot,\cdot)$ and $k_{d}(\cdot,\cdot)$ along with $\rho$ by maximizing 
the following log marginal likelihood: 
\begin{equation}
  \label{eq:lml2}
  \ln \wtl=-\dfrac{1}{2}(\wty-\wtm)^\trans\wtC^{-1} (\wty
  -\wtm)-\dfrac{1}{2}\ln \big\vert\wtC\big\vert-\dfrac{N_{H}+N_{L}}{2}\ln 2\pi.
\end{equation}

\subsection{GE-Kriging/Cokriging}
\label{subsec:gek_geck}
GE-Kriging uses the fact that under some condition, the derivative in physical space and the integral in the probability space are interchangable:
\begin{equation}
\label{eq:interchange}
\begin{aligned}
    \dfrac{\partial}{\partial x_i}\mu(\bm x) & =\dfrac{\partial}{\partial x_i} \mexp{Y(\bm x)}
    = \mexp{\dfrac{\partial}{\partial x_i} Y(\bm x)},\\ 
    \dfrac{\partial}{\partial x_i} k(\bm x,\bm x') & = \dfrac{\partial}{\partial x_i}\cov\left\{Y(\bm x), Y(\bm x')\right\} = \cov\left\{\dfrac{\partial}{\partial x_i}Y(\bm x),  Y(\bm x')\right\}, \\
   \dfrac{\partial^2}{\partial x_i\partial x'_j} k(\bm x,\bm x') & = \dfrac{\partial^2}{\partial x_i\partial x'_j}\cov\left\{Y(\bm x), Y(\bm x')\right\} = \cov\left\{\dfrac{\partial}{\partial x_i} Y(\bm x), \dfrac{\partial}{\partial x'_j} Y(\bm x')\right\}.
  \end{aligned}
  \end{equation}
These formulas specify the covariance between the QoI and its gradient as well as the covariance between different components of the gradient. To simplify the notations, we use $\partial_i$ and $\partial_{i'}$ to denote $\dfrac{\partial}{\partial x_i}$ and $\dfrac{\partial}{\partial x_i'}$, respectively, and $\nabla=(\partial_1, \partial_2, \dots, \partial_d )^\trans$, $\nabla'=(\partial_{1'}, \partial_{2'}, \dots, \partial_{d'})$. Of note, for a scalar function $z$, $\nabla z$ is a column vector and $\nabla' z$ is a row vector. Since we use a stationary kernel in this work, i.e., $k(\bm x, \bm x')=k(\bm x-\bm x')$, we have
\begin{equation}
   \partial_i k(\bm x,\bm x')=-\partial_{i'} k(\bm x,\bm x'). 
\end{equation}
The analytical form of $\partial_i k(\bm x, \bm x')$ and $\partial_i\partial_{j'} k(\bm x, \bm x')$  can be found in the appendix of~\cite{ulaganathan2015performance} for widely used kernel functions $k(\bm x, \bm x')$, e.g., Mat\'{e}rn kernels with several specific selections of $\nu$.  Subsequently, GE-Kriging follows almost the same procedures as those in Kriging with the following modifications~\cite{laurent2019overview}:
\begin{enumerate}
    \item The observation vector is augmented to include gradient data, i.e., 
    \begin{gather*}
    \bm y = (y^{(1)}, y^{(2)}, \dots, y^{(N)}, (\nabla y^{(1)})^\trans, (\nabla y^{(2)})^\trans, \dots, (\nabla y^{(N)})^\trans 
)^\trans.
    \end{gather*}
    \item Given a constant posterior mean of the QoI, the posterior mean of the gradient is zero, hence, $\bm 1 = (\underbrace{1, 1, \dots, 1}_{N}, \underbrace{0, 0, \dots, 0}_{N\times d})^\trans$.
    \item Covariance matrix $\tensor C = \sigma^2\tensor \Psi$, more specifically, the correlation matrix $\tensor \Psi$ is expanded to include correlations between QoI and its gradient as well as correlations between components of the gradient, i.e., 
    \begin{equation}
    \label{eq:gekrig_Psi}
    \tensor\Psi = \begin{bmatrix}
    \tensor\Psi_{11} & \tensor\Psi_{12}\\
    \tensor\Psi_{21} & \tensor\Psi_{22}
    \end{bmatrix},
    \end{equation}
    where
    \begin{align*}
   & \tensor\Psi_{11} = \dfrac{1}{\sigma^2}\begin{bmatrix}
    k(\bm x^{(1)}, \bm x^{(1)}) & \cdots & k(\bm x^{(1)}, \bm x^{(N)}) \\
    \vdots                      & \ddots & \vdots                      \\
    k(\bm x^{(N)}, \bm x^{(1)}) & \cdots & k(\bm x^{(N)},\bm x^{(N)})  
    \end{bmatrix},\\
    &\tensor\Psi_{21} = \nabla\tensor\Psi_{11}= \dfrac{1}{\sigma^2}
    \begin{bmatrix}
\partial_{1} k(\bm x^{(1)}, \bm x^{(1)})  &  \cdots & \partial_{1} k(\bm x^{(1)}, \bm x^{(N)}) \\
   \vdots & \ddots & \vdots \\ 
    \partial_{d} k(\bm x^{(1)}, \bm x^{(1)}) &  \cdots & 
    \partial_{d} k(\bm x^{(1)}, \bm x^{(N)}) \\
    \vdots & \ddots & \vdots \\
    \partial_{1} k(\bm x^{(N)}, \bm x^{(1)})  &  \cdots & \partial_{1} k(\bm x^{(N)}, \bm x^{(N)}) \\
   \vdots & \ddots & \vdots \\ 
    \partial_{d} k(\bm x^{(N)}, \bm x^{(1)}) &  \cdots & 
    \partial_{d} k(\bm x^{(N)}, \bm x^{(N)}) 
    \end{bmatrix}, \quad
    \tensor\Psi_{12} = \tensor\Psi_{21}^\trans, \\
    & \tensor\Psi_{22} = \nabla' \nabla\tensor\Psi_{11}=
    \begin{bmatrix}
    \bm\psi_{11} &  \cdots & \bm \psi_{1N} \\
    \vdots            & \ddots & \vdots\\
    \bm \psi_{N1} & \cdots & \bm \psi_{NN} 
    \end{bmatrix}, \quad
     \bm \psi_{lm} = \dfrac{1}{\sigma^2}
    \begin{bmatrix}
\partial_1\partial_{1'} k(\bm x^{(l)}, \bm x^{(m)})
 & \cdots &
\partial_1\partial_{d'} k(\bm x^{(l)}, \bm x^{(m)}) \\
\vdots& \ddots & \vdots\\
\partial_d\partial_{1'} k(\bm x^{(l)}, \bm x^{(m)})
 & \cdots &
\partial_d\partial_{d'} k(\bm x^{(l)}, \bm x^{(m)})
    \end{bmatrix}. 
    \end{align*}
\end{enumerate}
The posterior mean and variance of the QoI at a new location $\bm x^*$, denoted by $\hat y(\bm x^*)$ and $\hat s^2(\bm x^*)$, has the same form as in Kriging, i.e., Eqs.~\eqref{eq:krig_mean} and \eqref{eq:krig_var}, except that $\bm\psi=\begin{pmatrix}\bm \psi(\bm x^*)\\ \nabla \bm\psi(\bm x^*)\end{pmatrix}$, where
$\nabla\bm\psi(\bm x^*)=\dfrac{1}{\sigma^2}\begin{pmatrix}\nabla k(\bm x^{(1)},\bm x^*)\\ \vdots \\ \nabla k(\bm x^{(N)},\bm x^*)\end{pmatrix}$. 
Furthermore, the posterior mean and variance of the QoI's gradient at $\bm x^*$ are computed as
\begin{align}
\widehat{\partial_i y}(\bm x^*)& =(\partial_{i'}\bm\psi)^\trans \tensor \Psi^{-1} (\bm y - \bm 1 \hat{\mu}), \\
    \widehat{s_i}^2(\bm x^*)& =\hat{\sigma}^2\left[1 - (\partial_{i'}\bm \psi)^\trans \tensor \Psi^{-1} \partial_{i'} \bm \psi\right],
\end{align}
where $\partial_{i'}\bm\psi=\begin{pmatrix} \partial_{i'}\bm\psi(\bm x^*)\\ \partial_{i'}(\nabla \bm\psi(\bm x^*))\end{pmatrix}$ and $i=1,2,\dots,d$.

Next, we introduce the details of GE-Cokriging method, which also shares a similar construction procedure as Cokriging except for some modifications to incorporate gradient information. Such modifications are as follows:
\begin{enumerate}
    \item The observation vector is augmented to $\wty = \left(\bm y_{L}^{\trans}, \bm y_{H}^{\trans}, (\nabla \bm y_{L})^{\trans}, (\nabla \bm y_{H})^{\trans} \right)^{\trans}$ and is of length $N_L + N_H + (N_L + N_H)d$. 
    \item The covariance matrix of the observation data, $\wtC$ in Eq.~\eqref{eq:cokrig_cov}, is augmented to include gradient information as well, i.e.,
    \begin{equation}
     \wtC=
    \begin{pmatrix}
    \wtC_{11} & \wtC_{12} \\
    \wtC_{21} & \wtC_{22}
    \end{pmatrix}
    \end{equation}
    where $\wtC_{11}$ takes the form of covariance matrix in Cokriging, see Eq.~\eqref{eq:cokrig_cov}, and
    \begin{align*}
        & \wtC_{21} =
        \begin{bmatrix}
        \nabla \tensor C_{L}(\bm X_{L}, \bm X_{L}) & \rho \nabla \tensor C_{L}(\bm X_{L}, \bm X_{H}) \\
        \nabla \tensor C_{L} (\bm X_{H}, \bm X_{L}) & \rho^2 \nabla \tensor C_{L}(\bm X_{H}, \bm X_{H}) + \nabla \tensor C_{d}(\bm X_{H}, \bm X_{H})
        \end{bmatrix}, 
        & \wtC_{12} = \wtC_{21}^\trans, \\
        & \wtC_{22} =
        \begin{bmatrix}
        \nabla' \nabla \tensor C_{L}(\bm X_{L}, \bm X_{L}) & \rho \nabla' \nabla \tensor C_{L}(\bm X_{L}, \bm X_{H}) \\
        \rho \nabla' \nabla \tensor C_{L} (\bm X_{H}, \bm X_{L}) & \rho^2 \nabla' \nabla\tensor C_{L}(\bm X_{H}, \bm X_{H}) + \nabla' \nabla \tensor C_{d}(\bm X_{H}, \bm X_{H})
        \end{bmatrix}. & 
    \end{align*}
    Here $\nabla \tensor C_{L}(\bm X_{L}, \bm X_{L})$ is a matrix constructed by replacing each element in $\tensor C_{L}(\bm X_{L}, \bm X_{L})$, i.e., $[\tensor C_{L}(\bm X_{L}, \bm X_{L})]_{ij}$, with its gradient $(\partial_{1}[\tensor C_{L}(\bm X_{L}, \bm X_{L})]_{ij}, \dots, \partial_{d}[\tensor C_{L}(\bm X_{L}, \bm X_{L})]_{ij})^\trans$. Similarly, $\nabla \tensor C_{L}(\bm X_{L}, \bm X_{H})$, $\nabla \tensor C_{L}(\bm X_{H}, \bm X_{L})$, $\nabla \tensor C_{L}(\bm X_{H}, \bm X_{H})$ and $\nabla \tensor C_{d}(\bm X_{H}, \bm X_{H})$ are constructed by replacing elements in corresponding matrices in Eq.~\eqref{eq:cokrig_ele} with their gradients, respectively. The matrix $\nabla' \nabla \tensor C_{L}(\bm X_{L}, \bm X_{L})$ is constructed by replacing each element in $\tensor C_{L}(\bm X_{L}, \bm X_{L})$, i.e., $[\tensor C_{L}(\bm X_{L}, \bm X_{L})]_{ij}$, with the matrix
    \[\begin{pmatrix}\partial_1 \partial_{1'}[\tensor C_{L}(\bm X_{L}, \bm X_{L})]_{ij} & \cdots & \partial_1 \partial_{d'}[\tensor C_{L}(\bm X_{L}, \bm X_{L})]_{ij} \\
    \vdots & \ddots & \vdots \\
    \partial_d \partial_{1'}[\tensor C_{L}(\bm X_{L}, \bm X_{L})]_{ij} & \cdots & \partial_d \partial_{d'}[\tensor C_{L}(\bm X_{L}, \bm X_{L})]_{ij} 
    \end{pmatrix}.\]
    Other submatrices in $\wtC_{22}$ are constructed in the same manner.
    \item The posterior mean vector now becomes
    \begin{equation}
        \label{eq:gecokrig_mu}
        \wtm = \begin{pmatrix}\bm\mu_{L}\\ \bm\mu_{H} \\ \bm 0_L \\ \bm 0_H\end{pmatrix}=
        \begin{pmatrix}
        \big(\mu_{L}(\bm x_{_{L}}^{(1)}),\dotsc, 
        \mu_{L}(\bm x_{_{L}}^{(N_{L})})\big)^\trans \\
        \big(\mu_{H}(\bm x_{_{H}}^{(1)}), \dotsc, 
        \mu_{H}(\bm x_{_{H}}^{(N_{H})})\big)^\trans \\
        (\underbrace{0, \dotsc, 0}_{N_{L}\cdot d})^\trans \\
        (\underbrace{0, \dotsc, 0}_{N_{H}\cdot d})^\trans
        \end{pmatrix}.
    \end{equation}
    \item The covariance vector between the new observation location $\bm x^*$ and existing observation data $[\bm X_L, \bm X_H]$, denoted by $\wtc(\bm x^*)$, is given by
    \begin{equation}
        \label{eq:gecokrig_c}
        \wtc(\bm x^*) = 
        \begin{pmatrix}
        \rho\bm c_{L}(\bm x^*)\\ \bm c_{H}(\bm x^*) \\
        \rho\nabla  \bm c_{L}(\bm x^*)\\
         \nabla \bm c_{H}(\bm x^*)
        \end{pmatrix},
    \end{equation}
    where $\bm c_L(\bm x^*)=\big( k_{L}(\bm x_{L}^{(1)}, \bm x^*),\dotsc, k_{L}(\bm x_{L}^{(N_{L})}, \bm x^*)\big)^\trans$ and $\bm c_H(\bm x^*)=\big( k_{H}(\bm x_{H}^{(1)}, \bm x^*),\dotsc, k_{H}(\bm x_{H}^{(N_{H})}, \bm x^*)\big)^\trans$.
\end{enumerate}
The estimators for the mean and standard deviation of QoI at the new observation location $\bm x^*$ in GE-Cokriging follow Eqs.~\eqref{eq:cokrig_mean} and~\eqref{eq:cokrig_var} in Cokriging method with corresponding components updated as shown above.

We provide the formulas for the posterior mean and variance of the QoI's gradient at $\bm x^*$ as follows:
\begin{align}
\label{eq:gecokrig_mean}
    \widehat{\partial_i y}(\bm x^*) & = (\partial_{i'} \wtc(\bm x^*))^\trans\wtC^{-1}(\wty-\wtm),\\
    \label{eq:gecokrig_var}
    \widehat{s_i}^2(\bm x^*) & =\rho^2 \partial_i \partial_{i'} k_L(\bm x^*, \bm x^*) + \partial_i \partial_{i'}k_H(\bm x^*, \bm x^*) - \left[\partial_{i'}\wtc(\bm x^*)\right]^\trans \wtC^{-1} \partial_{i'} \wtc(\bm x^*),
\end{align}
where $i=1,2,\dotsc, d$. The derivation of Eqs.~\eqref{eq:gecokrig_mean} and~\eqref{eq:gecokrig_var} follow the same procedure as Eqs.~\eqref{eq:cokrig_mean} and~\eqref{eq:cokrig_var} shown in~\cite{kennedy2000predicting, forrester2008engineering}. In other words, Eqs.~\eqref{eq:gecokrig_mean} and~\eqref{eq:gecokrig_var} can be obtained by replacing $Y(\bm x)$ in Eqs.~\eqref{eq:cokrig_mean} and~\eqref{eq:cokrig_var} with $\partial_i Y(\bm x)$. More specifically, $\mu_H(\bm x^*)$ is replaced with the mean of $\partial_i Y(\bm x)$ (which is zero), $\wtc$ is replaced with $\partial_i \wtc$, and $\rho^2\sigma_L^2(\bm x^*)+\sigma_d^2(\bm x^*)$ (i.e., $\rho^2\text{Var}\{Y_L(\bm x^*)\}+\text{Var}\{Y_d(\bm x^*) \}$) is replaced with $\rho^2\text{Var}\{\partial_i Y_L(\bm x^*)\}+\text{Var}\{\partial_i Y_d(\bm x^*) \}=\rho^2 \partial_i \partial_{i'} k_L(\bm x^*, \bm x^*) + \partial_i \partial_{i'}k_H(\bm x^*, \bm x^*)$.

We note that the GE-Cokriging exploits the relation between QoI and its gradients, and once the hyperparameters in the model are identified, we can compute the posterior mean and variance of the QoI and its gradients \emph{simultaneously}. It has the potential to improve the accuracy of the prediction for both QoI and its gradients compared with predicting them separately. Also, in some cases, this approach can reduce computational cost compared to, for example, constructing Cokriging models for QoI and its gradients separately (see Section~\ref{subsec:err}).

\subsection{Integral-enhanced Kriging/Cokriging}
\label{subsec:integral}
In this section, we provide another perspective on using the QoI $f$ and its gradients $\nabla f$ in GPR simultaneously. The aforementioned gradient-enhanced methods firstly assume a GP model $Y(\bm x)$ for $f$, and the GP model for $\nabla f$ can be constructed accordingly by taking (partial) derivatives of $Y(\bm x)$'s mean and covariance function. Alternatively, one can also assume a GP model for $\nabla f$ first, e.g., $\partial_i f$ is modeled by $Y(\bm x)$, then the QoI $f$ can be modeled by $\int Y(\bm x)\dif x_i$, which is a GP because integral is a linear operator. Here we use the univariate function to further illustrate the concept. We model $f'$ with GP $Y_{f'}(\bm x)\sim\mathcal{GP}(\mu_{f'}(\bm x), k_{f'}(\bm x, \bm x'))$, then similar to Eqs.~\eqref{eq:interchange}, the integrals in the physical space and in the probability space are interchangeable:
\begin{equation}
\label{eq:integral}
\begin{aligned}
    \int \mu_{f'}(\bm x) \dif \bm x & =\int \mexp{Y_{f'}(\bm x)} \dif \bm x
    = \mexp{\int Y_{f'}(\bm x) \dif \bm x},\\ 
    \int k_{f'}(\bm x, \bm x') \dif \bm x & = \int\cov\left\{Y_{f'}(\bm x), Y_{f'}(\bm x')\right\}\dif \bm x \\
    &=\int \mexp{(Y_{f'}(\bm x)-\mu_{f'}(\bm x))(Y_{f'}(\bm x')-\mu_{f'}(\bm x'))}\dif \bm x \\
    &= \mexp{\bigg[\int (Y_{f'}(\bm x)-\mu_{f'}(\bm x))\dif \bm x \bigg] (Y_{f'}(\bm x')-\mu_{f'}(\bm x'))} \\
    &=\cov\left\{\int Y_{f'}(\bm x) \dif \bm x,  Y_{f'}(\bm x')\right\}, \\
   \int \int k_{f'}(\bm x, \bm x') \dif \bm x \dif \bm x'& = \int \int\cov\left\{Y_{f'}(\bm x), Y_{f'}(\bm x')\right\} \dif \bm x \dif \bm x' \\ &
   = \int \int\mexp{(Y_{f'}(\bm x)-\mu_{f'}(\bm x))(Y_{f'}(\bm x')-\mu_{f'}(\bm x'))} \dif \bm x \dif \bm x'\\
   &= \mexp{\int (Y_{f'}(\bm x)-\mu_{f'}(\bm x)) \dif \bm x \int(Y_{f'}(\bm x')-\mu_{f'}(\bm x'))\dif \bm x'}\\
   & = \cov\left\{\int Y_{f'}(\bm x) \dif \bm x, \int Y_{f'}(\bm x')\dif \bm x'\right\}.
  \end{aligned}
  \end{equation}
These formulas provide the mean and covariance of the GP $Y_f(\bm x)=\int Y_{f'}(\bm x)\dif \bm x$ as well as the covariance between $Y_f(\bm x)$ and $Y_{f'}(\bm x)$.
Of note, we use indefinite integral here and the constant associated with this integral needs identification via maximizing the log marginal likelihood. But this constant will not affect the covariance function, because $\cov\left\{\int Y_{f'}(\bm x) \dif \bm x, \int Y_{f'}(\bm x')\dif \bm x'\right\}=\cov\left\{\int Y_{f'}(\bm x) \dif \bm x+a \int Y_{f'}(\bm x')\dif \bm x'+b\right\}$ for any constants $a$ and $b$.  
  
Then we can follow the same procedure in the gradient-enhanced Kriging in Section~\ref{subsec:gek_geck} to construct the covariance matrix $\tensor C$ and compute the posterior mean and variance of $f$ and $f'$ at any location $\bm x^*$.
Of note, this ``integral-enhanced'' GPR/Kriging is equivalent to the gradient-enhanced version. For example, if we set the mean of $Y_{f'}(\bm x)$ to be zero, then the mean of $Y_f(\bm x)$ is a constant $\mu$, which needs identifying as in the gradient-enhanced version. 
Subsequently, the integral-enhanced Kriging is equivalent to the equivalence of the gradient-enhanced Kriging if the mean and covariance functions are selected appropriately.
For example, if we assume zero mean and set $k_{f'}(\bm x , \bm x')=\dfrac{\partial^2}{\partial x_i\partial x_j'}k_f(\bm x, \bm x')$ for $Y_{f'}(\bm x)$, where $k_f(\bm x, \bm x')$ is the Gaussian kernel function, this integral-enhanced Kriging model is the same as the gradient-enhanced Kriging model that uses Gaussian kernel function and constant mean for $Y_f(\bm x)$. In most cases, it is easier to compute the (partial) derivatives than to compute the integral. Therefore, it is more convenient to use the gradient-enhanced setting. The similar argument holds for Cokriging. In this work, we only show the results of gradient-enhanced Kriging/Cokriging.

\section{Numerical examples}
\label{sec:numeric}

We present four numerical examples to demonstrate the performance of GE-Cokriging. The first two prototype examples show the capability GE-Cokriging's capability of approximating the QoI and its gradients of two 1D functions and a 2D function. The other two examples illustrate the high precision of GE-Cokriging in
constructing the phase diagram of an underdamped oscillator and analyzing the sensitivity of power factor under varying power inputs in a large-scale power grid system. In all these examples, we assume that both the QoI and its gradients are collected at every observation locations. The hyperparameters in GP models are identified by maximizing associated log marginal likelihood function using genetic algorithm as in~\cite{forrester2008engineering}. Lastly, we compare the prediction accuracy using Cokriging, GE-Kriging and GE-Cokriging in each case quantitatively. We also compare the computational cost of these methods in each case.

\subsection{1D function}

In this part, we compare the results of Cokriging and GE-Cokriging in approximating a 1D function. In this case, the target function to approximate is,
\begin{equation}
f_H(x) = (6x-2)^2\sin(12x-4),
\end{equation}
from which high-fidelity data are sampled. The low-fidelity data are sampled from the following function
\begin{equation}
f_L(x) = Af_H(x) + B(x-0.5) + C.
\end{equation}
The observation locations of 
$f_H$ are $X_H = \{0, 0.2, 0.6, 1.0\}$, and those for 
$f_L$ are $X_L = \{0,  0.2, 0.4,  0.6, 0.8, 1.0\}$. Here, the observation locations of data are chosen so that $X_H \subset X_L$. 

\subsubsection{1D Case 1: a classical case\label{subsec:1d1}}
We first show a well-studied case where parameters of low-fidelity function is given by $A = 0.5, B = 10, C = -5$ as in~\cite{forrester2008engineering}. Hence, the low-fidelity function is
\begin{equation}
    \label{eq:fc1}
    f_{L1}(x) = 0.5f_H(x) + 10(x-0.5) - 5.
\end{equation}
Of note, we use fewer observation points in $\bm X_L$ than in~\cite{forrester2008engineering}.

\begin{figure}[h!]
    \centering
    \begin{subfigure}{0.49\textwidth}
        \includegraphics[width=\textwidth]{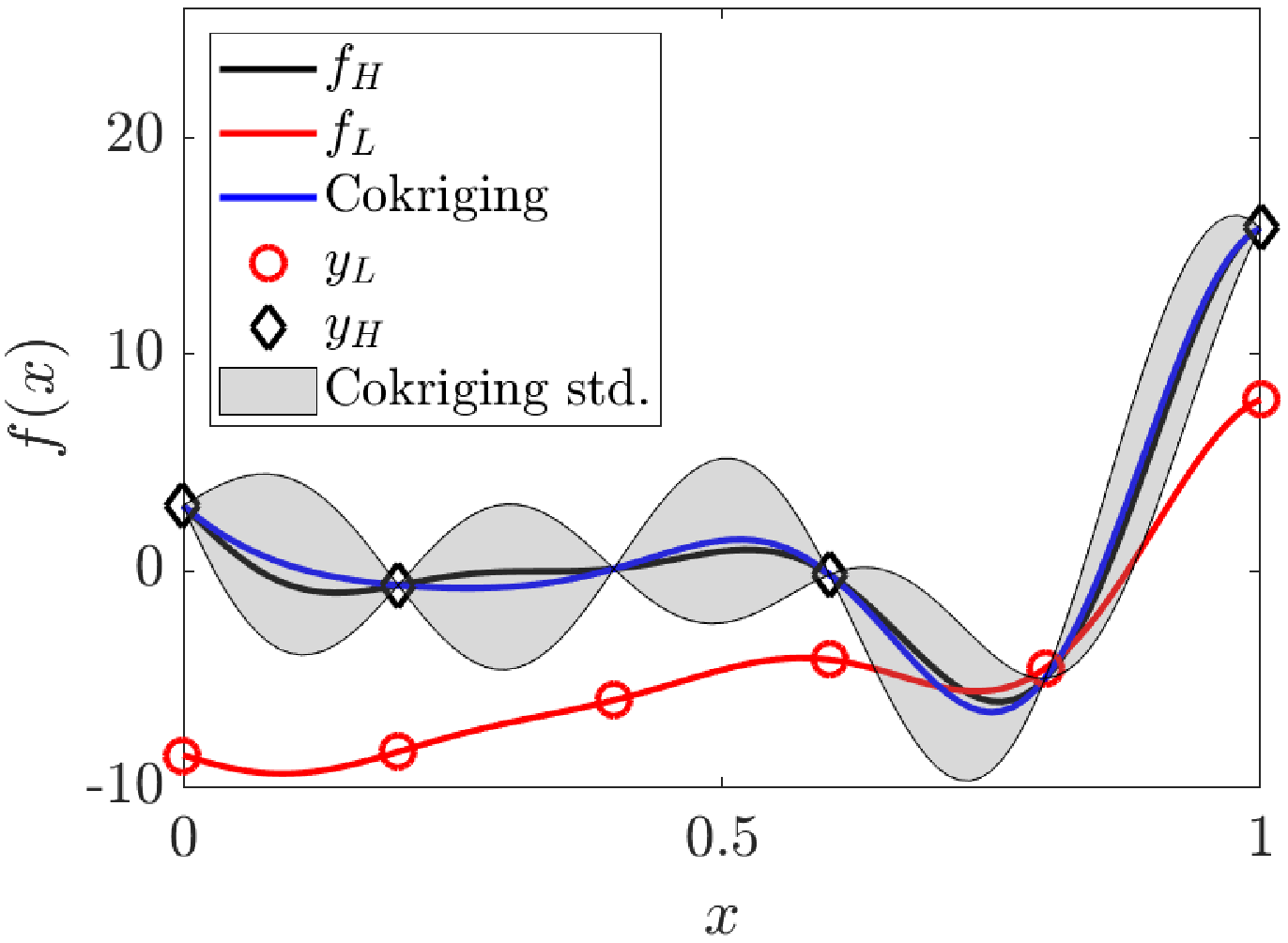}
        \caption{}
        \label{fig:1d1_val_cokrig}
    \end{subfigure}
    \begin{subfigure}{0.49\textwidth}
        \includegraphics[width=\textwidth]{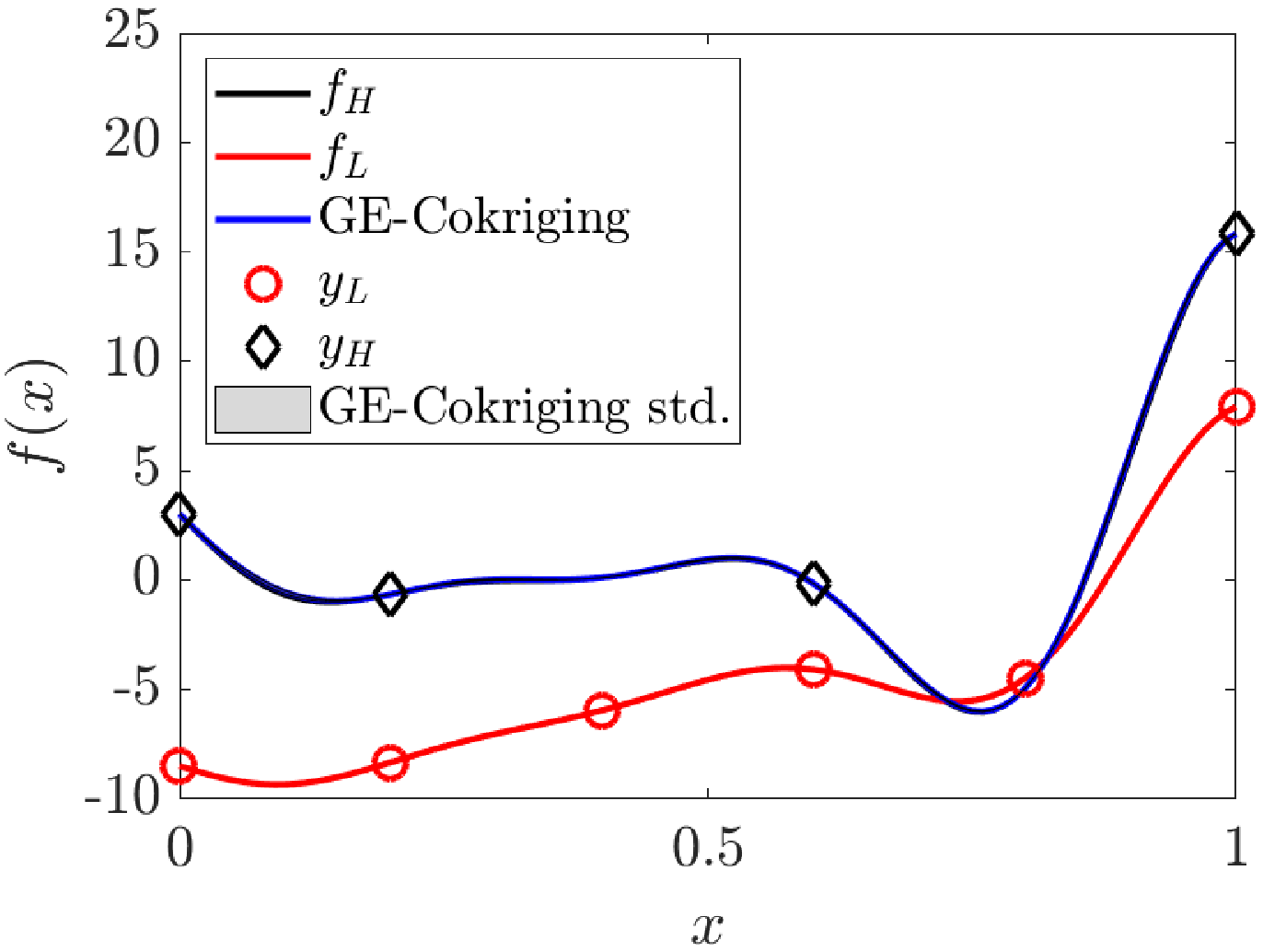}
        \caption{}
        \label{fig:1d1_val_geck}
    \end{subfigure}
    \caption{Prediction of the QoI for the 1D problem case 1. Prediction of posterior mean (black solid line) and standard deviation (grey shaded area) of QoI $f_H$  by (a) Cokriging and (b) GE-Cokriging. The low-fidelity function $f_{L1}$ is denoted by red solid lines, high-fidelity samples are denoted by black diamonds and low-fidelity samples by red circles. Colored online.}
    \label{fig:1d1_val}
\end{figure}
The results of Cokriging and GE-Cokriging 
for reconstructing $f_H$
are shown in Fig.~\ref{fig:1d1_val}. 
Fig.~\ref{fig:1d1_val_cokrig} shows that Cokriging is able to capture 
$f_H$ as the posterior mean is generally close to the high-fidelity function value. However, $\hat s$ of the prediction are large on most of the prediction locations, which indicates that Cokriging method yields considerable uncertainty at those locations, whereas 
this uncertainty is very small at $X_c$ because a simple relation has been found between $f_H$ and $f_L$ based on available data~\cite{forrester2008engineering}.
As a comparison, Fig.~\ref{fig:1d1_val_geck} illustrates that the posterior mean of GE-Cokriging coincides with $f_H$, and the uncertainty in the prediction is very small on the entire interval as the grey shaded area is almost invisible.

\begin{figure}[h!]
    \centering
    \begin{subfigure}{0.45\textwidth}
        \centering
        \includegraphics[width=\textwidth]{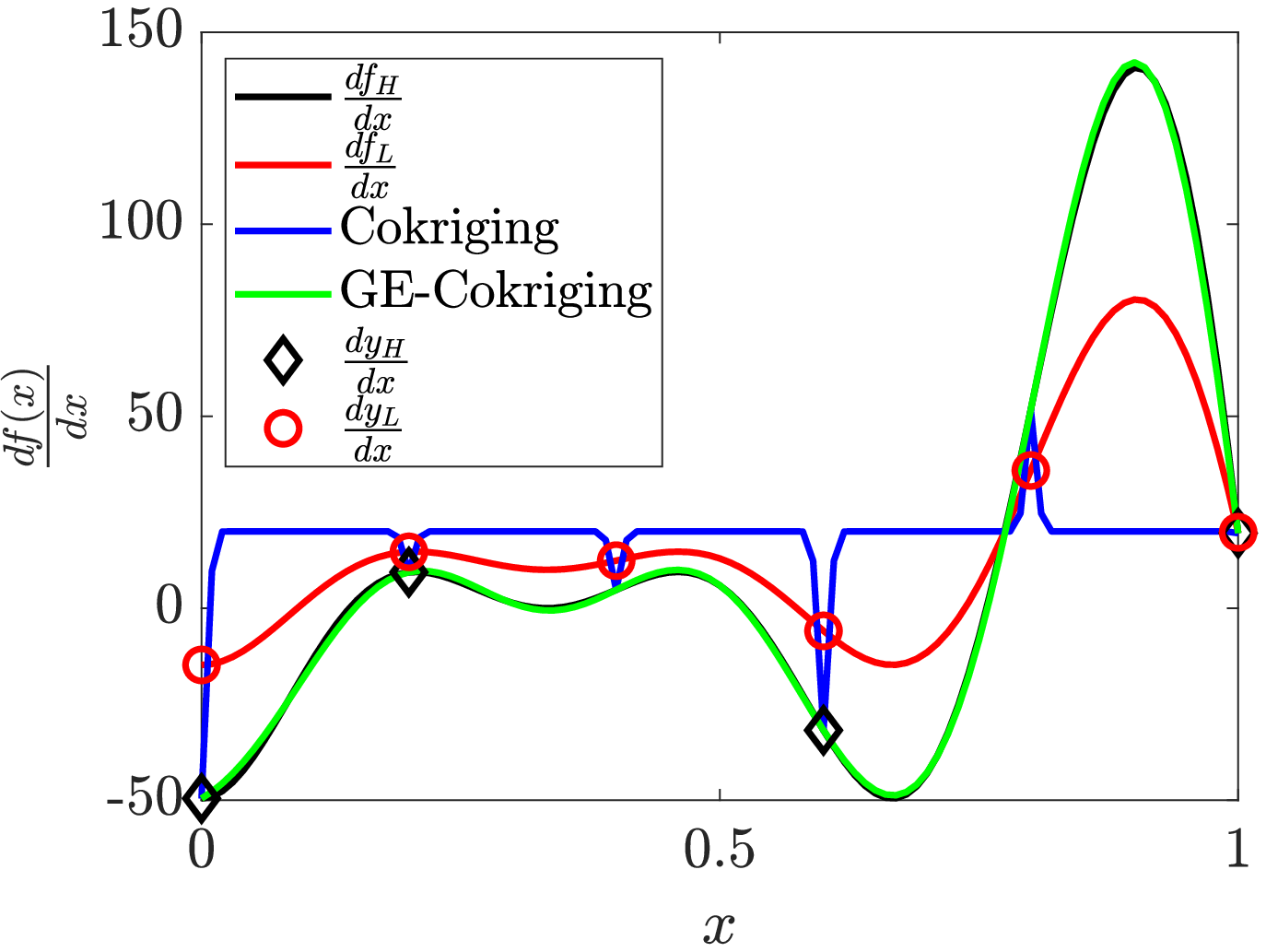}
        \caption{}
        \label{fig:1d1_grad_mean}
    \end{subfigure}
    \hspace{0.04\textwidth}
    \begin{subfigure}{0.45\textwidth}
        \centering
        \includegraphics[width=\textwidth]{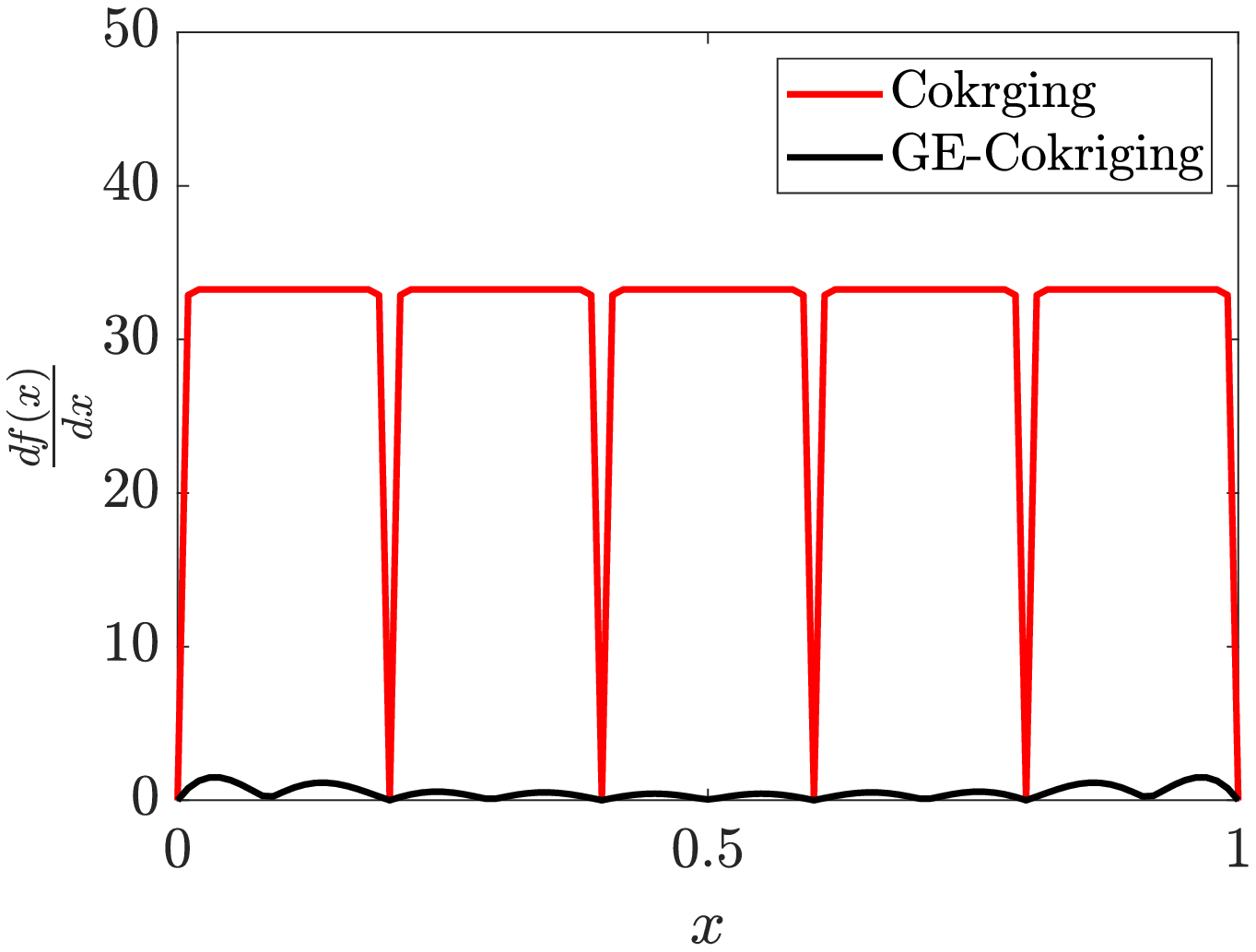}
        \caption{}
        \label{fig:1d1_grad_std}
    \end{subfigure}
    \caption{Prediction of the gradient of QoI for the 1D problem case 1. Prediction of posterior (a) mean by Cokriging (blue solid line) and GE-Cokriging (green solid line), where the gradient of high-fidelity function {$\frac{\dif f_H}{\dif x}$} is denoted by black solid line, gradient of low-fidelity function {$\frac{\dif f_{L1}}{\dif x}$} is denoted by red solid line, high-fidelity samples are denoted by black diamonds and low-fidelity samples by red circles and (b) standard deviation for gradient of QoI $\frac{\dif f_H}{\dif x}$ by Cokriging (red solid line) and GE-Cokriging (black solid line). Colored online.}
    \label{fig:1d1_grad}
\end{figure}

Next, we compare the performance of predicting the gradients of $f_H$, i.e., $\frac{\dif f_H(x)}{\dif x}$. Fig.~\ref{fig:1d1_grad} shows that Cokriging method suffers from the singularity of the covariance matrix in this setup, implied from sharp turning of predicted curvature between neighboring observations in Fig.~\ref{fig:1d1_grad_mean} and large standard deviations in Fig.~\ref{fig:1d1_grad_std} on locations where observations are not available. As for GE-Cokriging method, the prediction of gradients is accurate both in terms of posterior mean illustrated in Fig.~\ref{fig:1d1_grad_mean} and standard deviation illustrated Fig.~\ref{fig:1d1_grad_std}, which shows that the prediction uncertainty by Cokriging is almost 10 times greater than that by GE-Cokriging. 
We note that 
the performance of Cokriging is poor in this case 
because the covariance matrix $\tilde{\tensor C}$ is close to a singular matrix. The reason for this phenomenon is that the value of $\frac{\dif y_L}{\dif x}$ is close at $x=0.2$ and $x=0.4$, as well as at $x=0$ and $x=0.6$. As we point out in Section~\ref{subsec:gpr}, this singularity issue is common for GPR method in practice, and the typical approach to alleviate this is to add a diagonal matrix $\alpha I$ to the covariance matrix, which is equivalent to add noises in the collected data. In this paper, we set $\alpha=10^{-14}$, which is much smaller than typical numbers used in practice, to demonstrate that the GE-Cokriging can help to alleviate the singularity issue without sacrificing accuracy of matching observation data.

\subsubsection{1D Case 2: shifted $f_{L1}$}
\begin{figure}[t!]
    \centering
    \begin{subfigure}{0.49\textwidth}
    \centering
        \includegraphics[width=\textwidth]{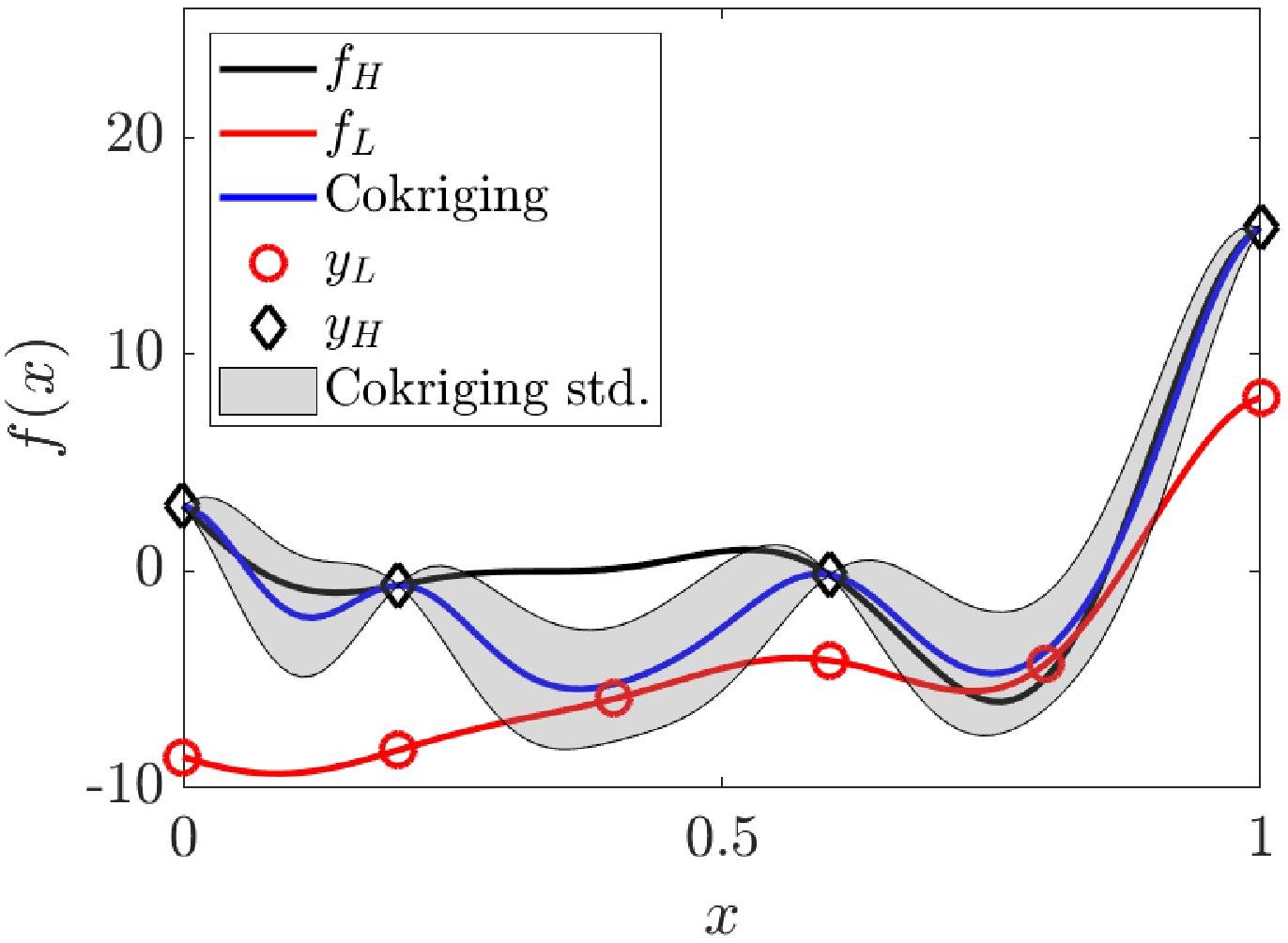}
        \caption{}
        \label{fig:1d2_val_cokrig}
    \end{subfigure}
    \begin{subfigure}{0.49\textwidth}
    \centering
        \includegraphics[width=\textwidth]{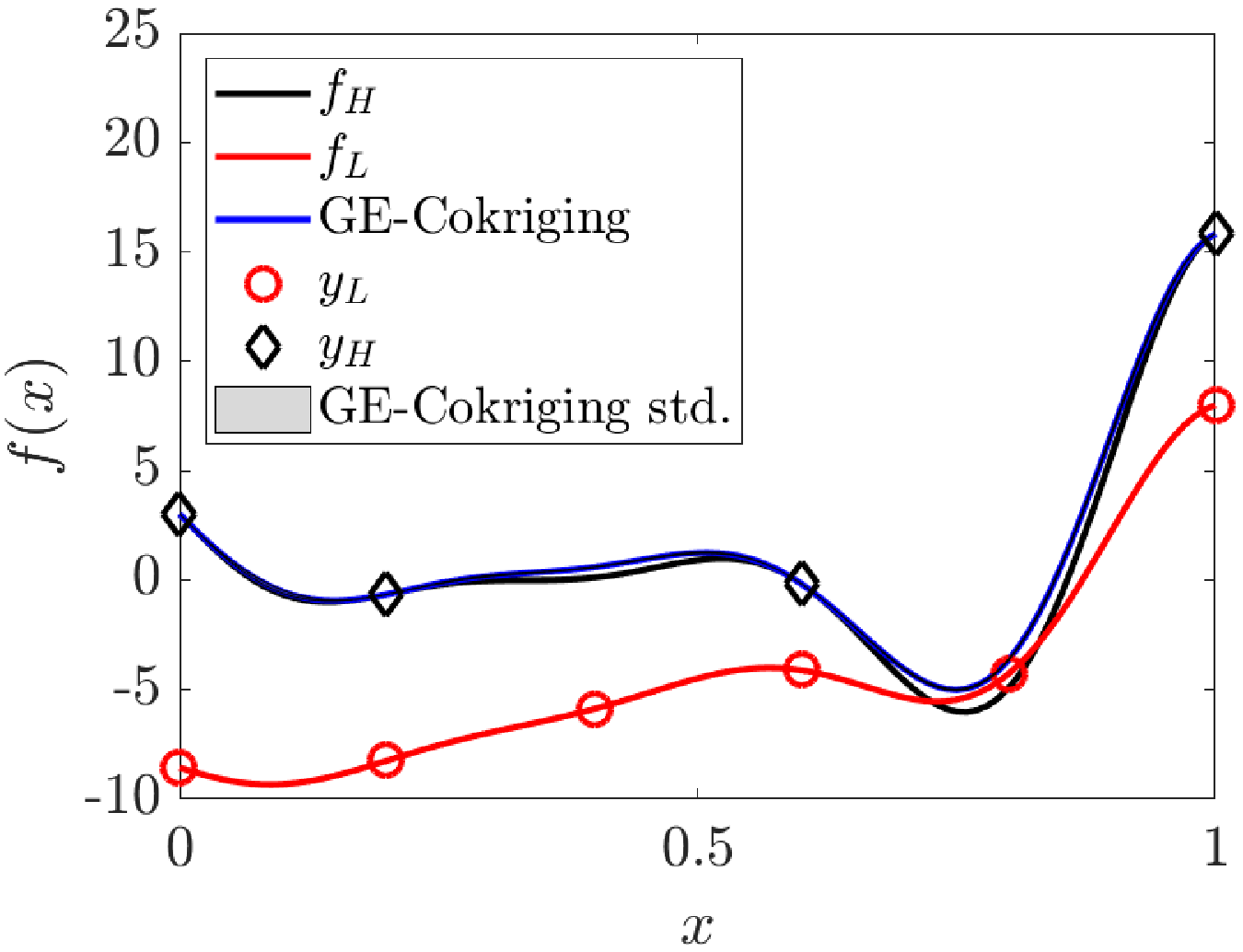}
        \caption{}
        \label{fig:1d2_val_geck}
    \end{subfigure}
    \caption{Prediction of the QoI for the 1D problem case 2. Prediction of posterior mean (black solid line) and standard deviation (grey shaded area) of QoI $f_H$  by (a) Cokriging and (b) GE-Cokriging. The low-fidelity function $f_{L2}$ is denoted by red solid lines, high-fidelity samples are denoted by black diamonds and low-fidelity samples by red circles. Colored online.}
    \label{fig:1d2_val}
\end{figure}

Next, we keep the sampling locations, i.e., $X_H$ and $X_L$ same as those in Section~\ref{subsec:1d1}, and only modify the model parameters of the low-fidelity function in Eq.~(\ref{eq:fc1}) by slightly shifting it, i.e., replace $x$ with $x-0.005$, resulting in the following form of low-fidelity function $f_{L2}$,
\begin{equation}
    f_{L2}(x) = f_{L1}(x - 0.005) = 0.5f_H(x - 0.005) + 10(x -0.005 - 0.5) - 5.
\end{equation}
The posterior means and standard deviations of Cokriging and GE-Cokriging are shown in Fig.~\ref{fig:1d2_val}. It is shown in Fig.~\ref{fig:1d2_val_cokrig} that the Cokriging method is not able to obtain an accurate prediction of $f_H$, and the resulting uncertainty is large on the entire interval except for locations of $X_H$. On the contrary, as shown in Fig.~\ref{fig:1d2_val_geck}, the GE-Cokriging result is much closer to $f_H$ and the uncertainty is very small.

\begin{figure}[!h]
    \centering
    \begin{subfigure}{0.45\textwidth}
        \includegraphics[width=\textwidth]{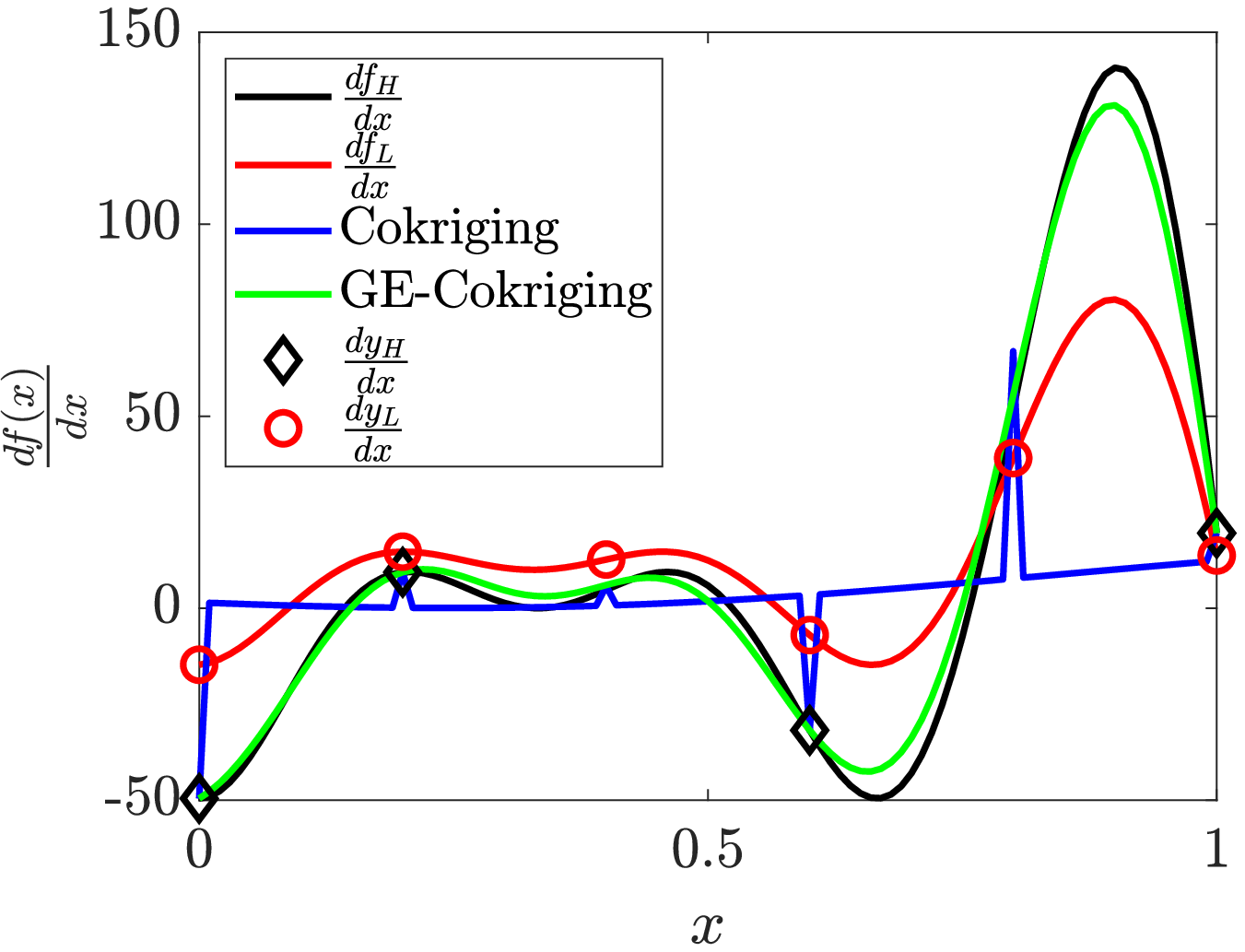}
        \caption{}
        \label{fig:1d2_grad_mean}
    \end{subfigure}
    \hspace{0.04\textwidth}
    \begin{subfigure}{0.45\textwidth}
        \includegraphics[width=\textwidth]{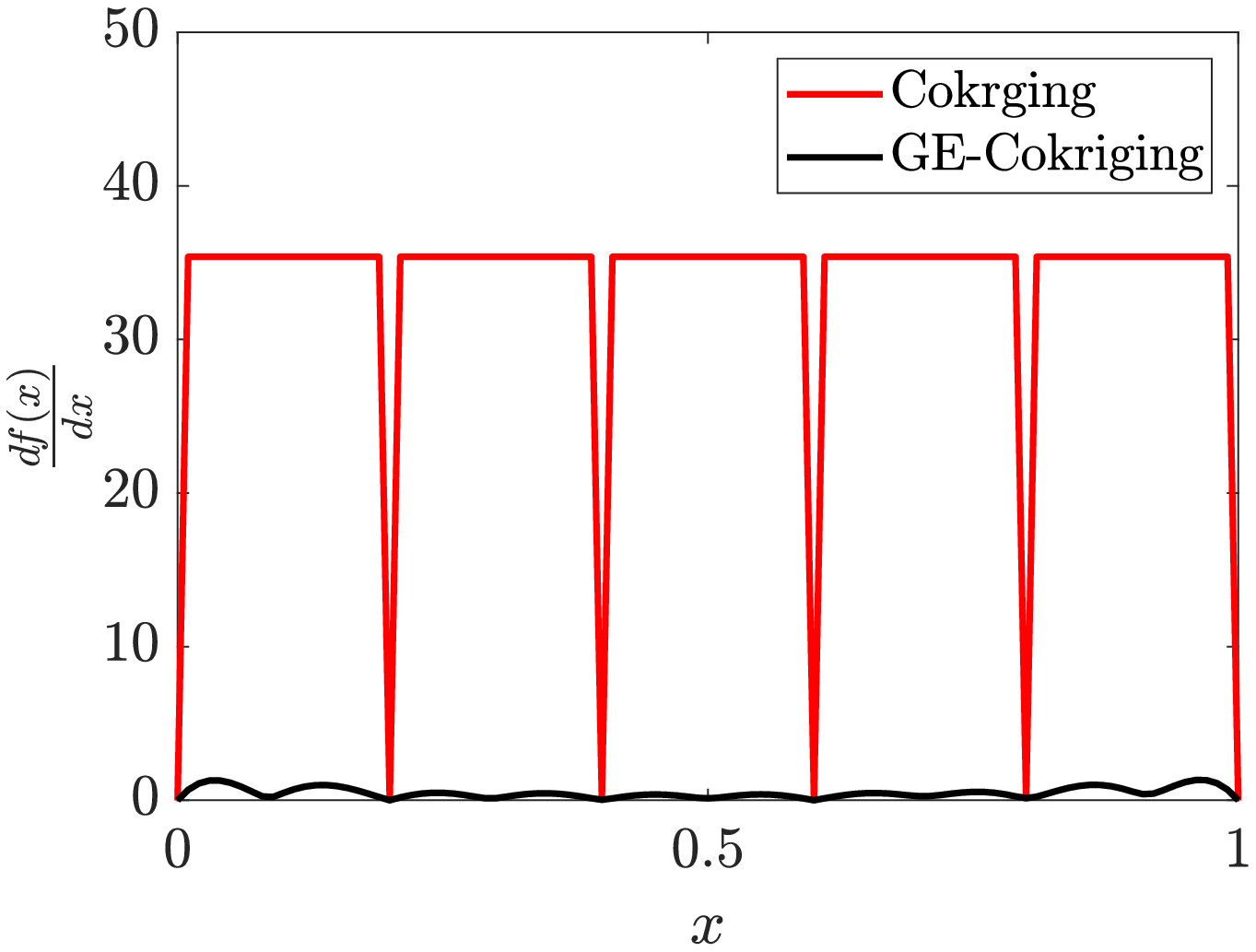}
        \caption{}
        \label{fig:1d2_grad_std}
    \end{subfigure}
    \caption{Prediction of the gradient of QoI for the 1D problem case 2. Prediction of posterior (a) mean by Cokriging (blue solid line) and GE-Cokriging (green solid line), where the gradient of high-fidelity function {$\frac{\dif f_H}{\dif x}$} is denoted by black solid line, gradient of low-fidelity function {$\frac{\dif f_{L2}}{\dif x}$} is denoted by red solid line, high-fidelity samples are denoted by black diamonds and low-fidelity samples by red circles and (b) standard deviation for gradient of QoI $\frac{\dif f_H}{\dif x}$ by Cokriging (red solid line) and GE-Cokriging (black solid line). Colored online.}
    \label{fig:1d2_grad}
\end{figure}

We present the prediction results of 
gradients by GE-Cokriging and Cokriging in Fig.~\ref{fig:1d2_grad}. Similar to the observations from Fig.~\ref{fig:1d1_grad_mean}, Cokriging in this case suffers from the singularity of the covariance matrix, with posterior mean deviating significantly from $f_H$ (see Fig.~\ref{fig:1d2_grad_mean}) and standard deviation being in the order comparable to its mean value (see Fig.~\ref{fig:1d2_grad_std}). In comparison, GE-Cokriging still yields a good result with posterior mean close to $\frac{\dif f_H}{\dif x}$ (see Fig.~\ref{fig:1d2_grad_mean}) and low uncertainty, i.e., small standard deviations (see Fig.~\ref{fig:1d2_grad_std}). These contrasts between the Cokriging and GE-Cokriging suggest that the gradient information from high-fidelity function and low-fidelity function can help to improve the prediction accuracy of not only QoI but also the corresponding gradients. 

\subsection{Branin function}
\begin{figure}[!h]
    \centering
    \begin{subfigure}{0.32\textwidth}
        \includegraphics[width=\textwidth]{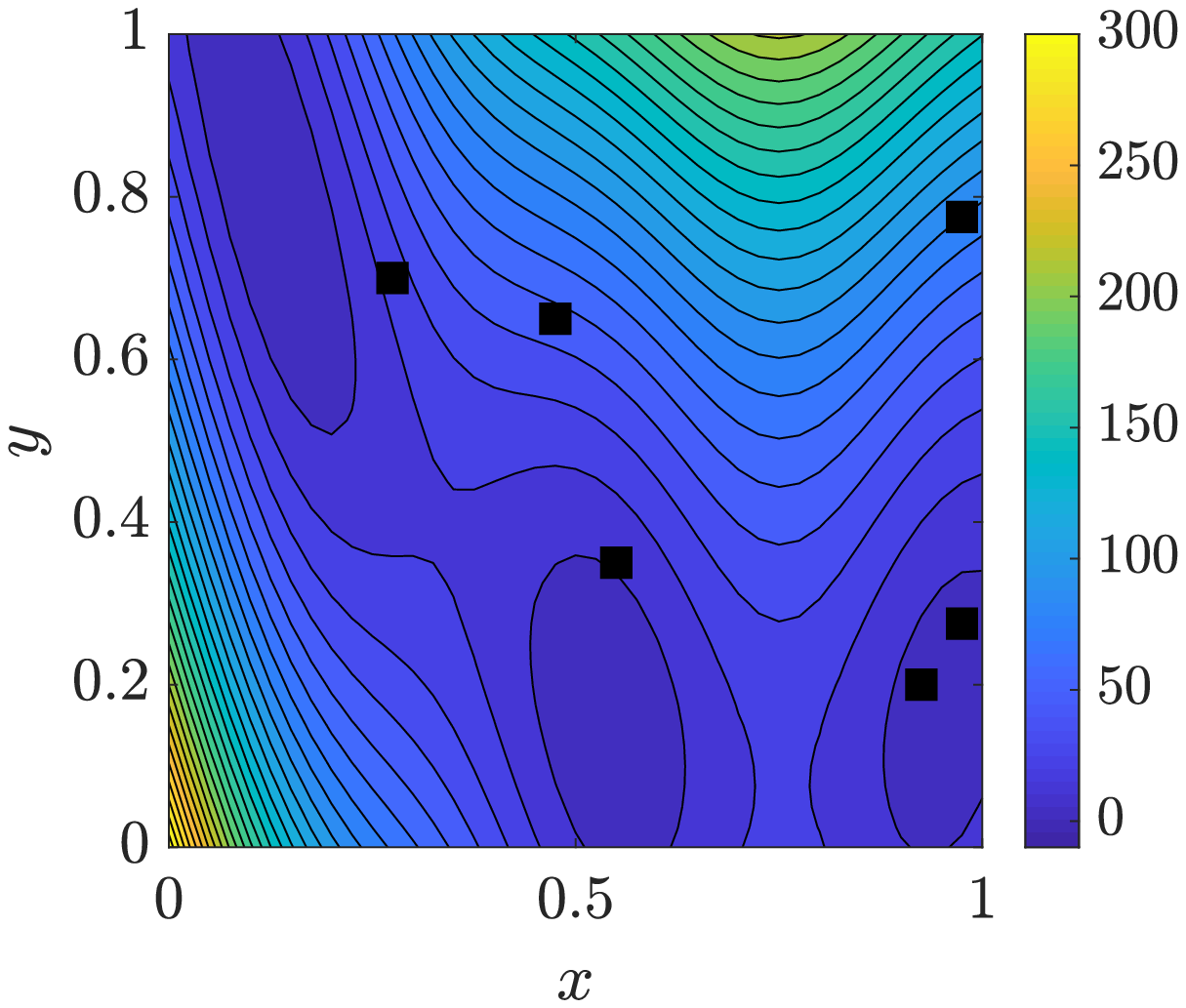}
        \caption{}
        \label{fig:2d_val_fe}
    \end{subfigure}
    \begin{subfigure}{0.32\textwidth}
        \includegraphics[width=\textwidth]{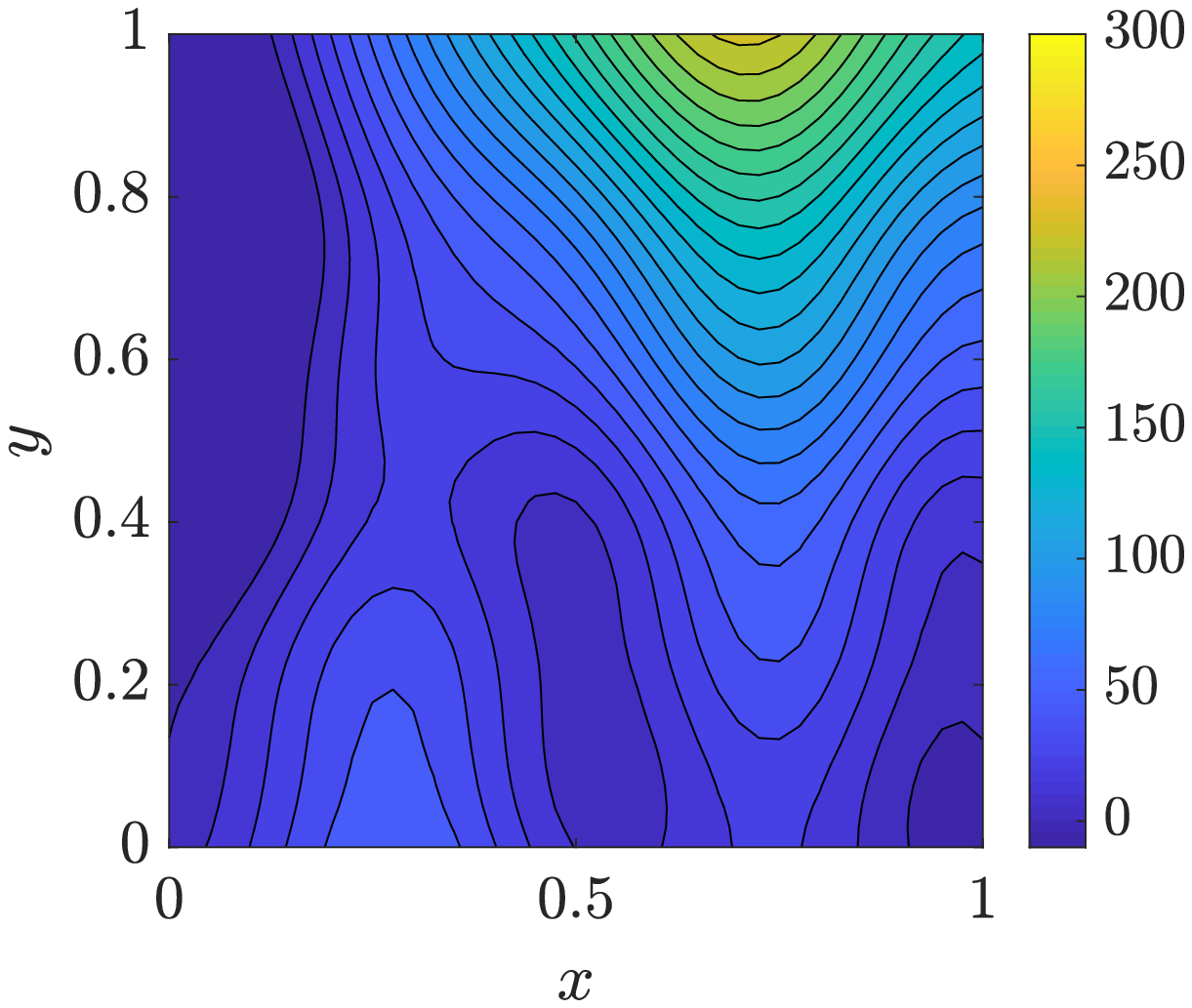}
        \caption{}
        \label{fig:2d_val_cokrig}
    \end{subfigure}
        \begin{subfigure}{0.32\textwidth}
        \includegraphics[width=\textwidth]{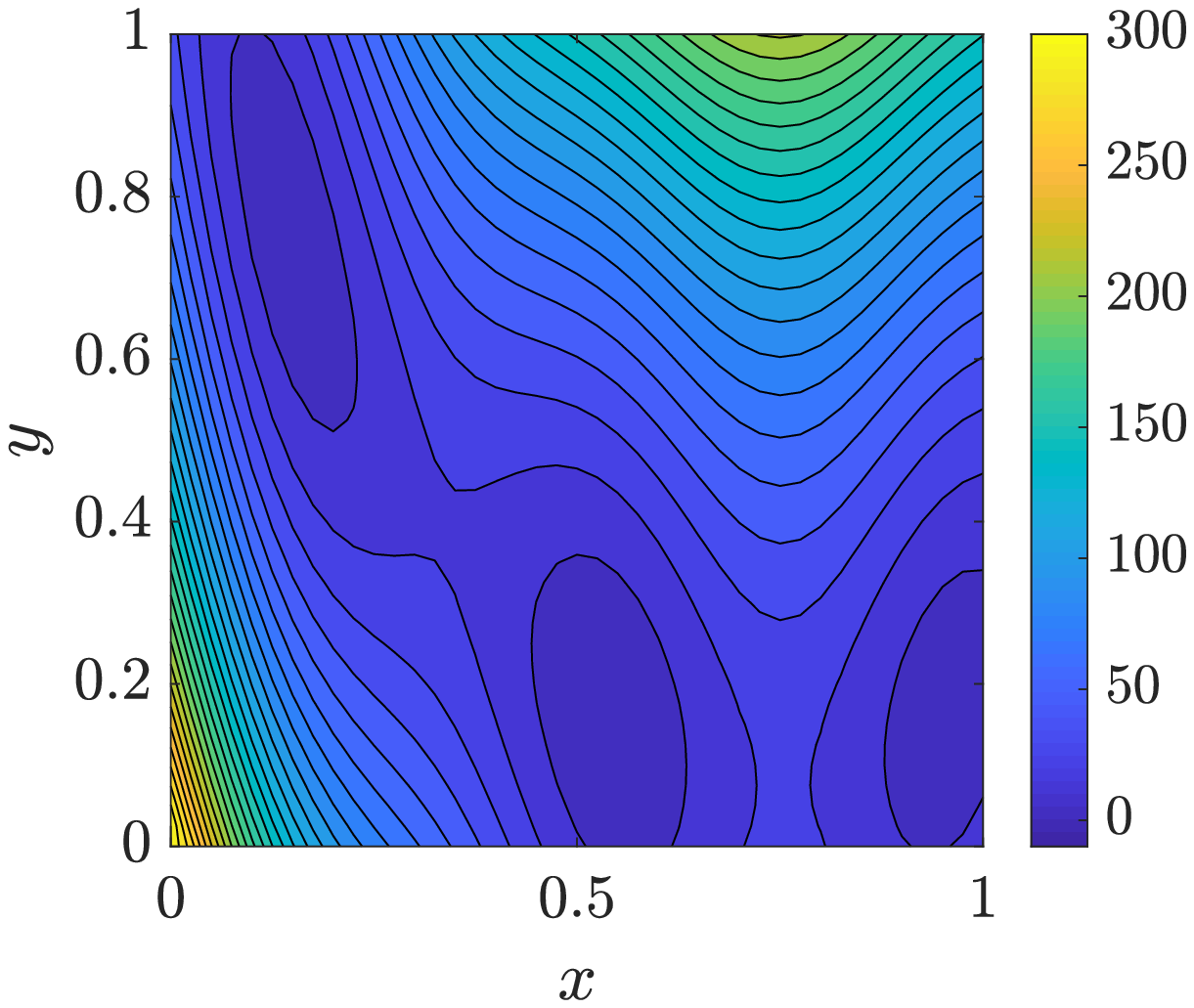}
        \caption{}
        \label{fig:2d_val_geck}
    \end{subfigure}
    ~
    \begin{subfigure}{0.32\textwidth}
        \includegraphics[width=\textwidth]{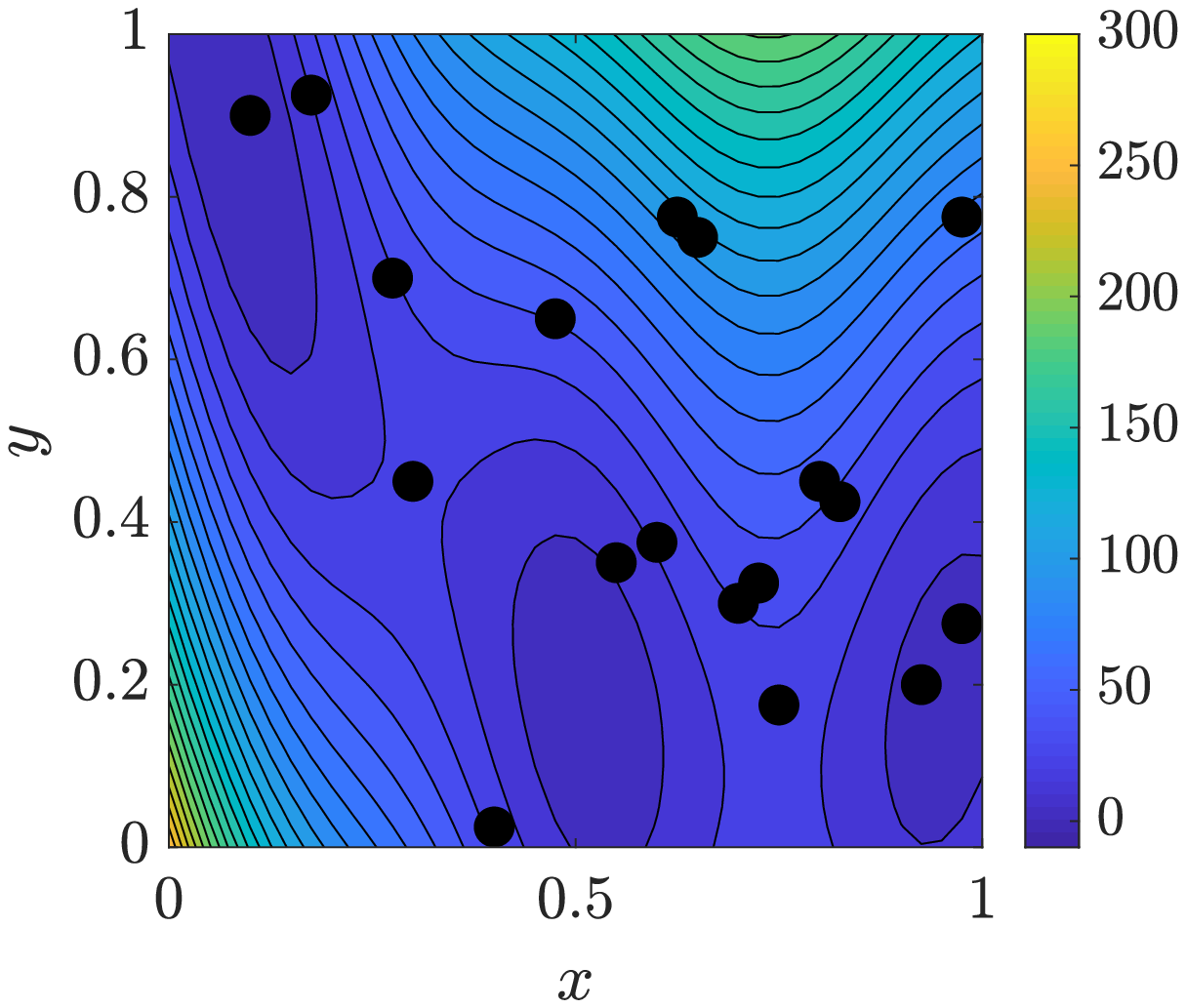}
        \caption{}
        \label{fig:2d_val_fc}
    \end{subfigure}
    \begin{subfigure}{0.32\textwidth}
        \includegraphics[width=\textwidth]{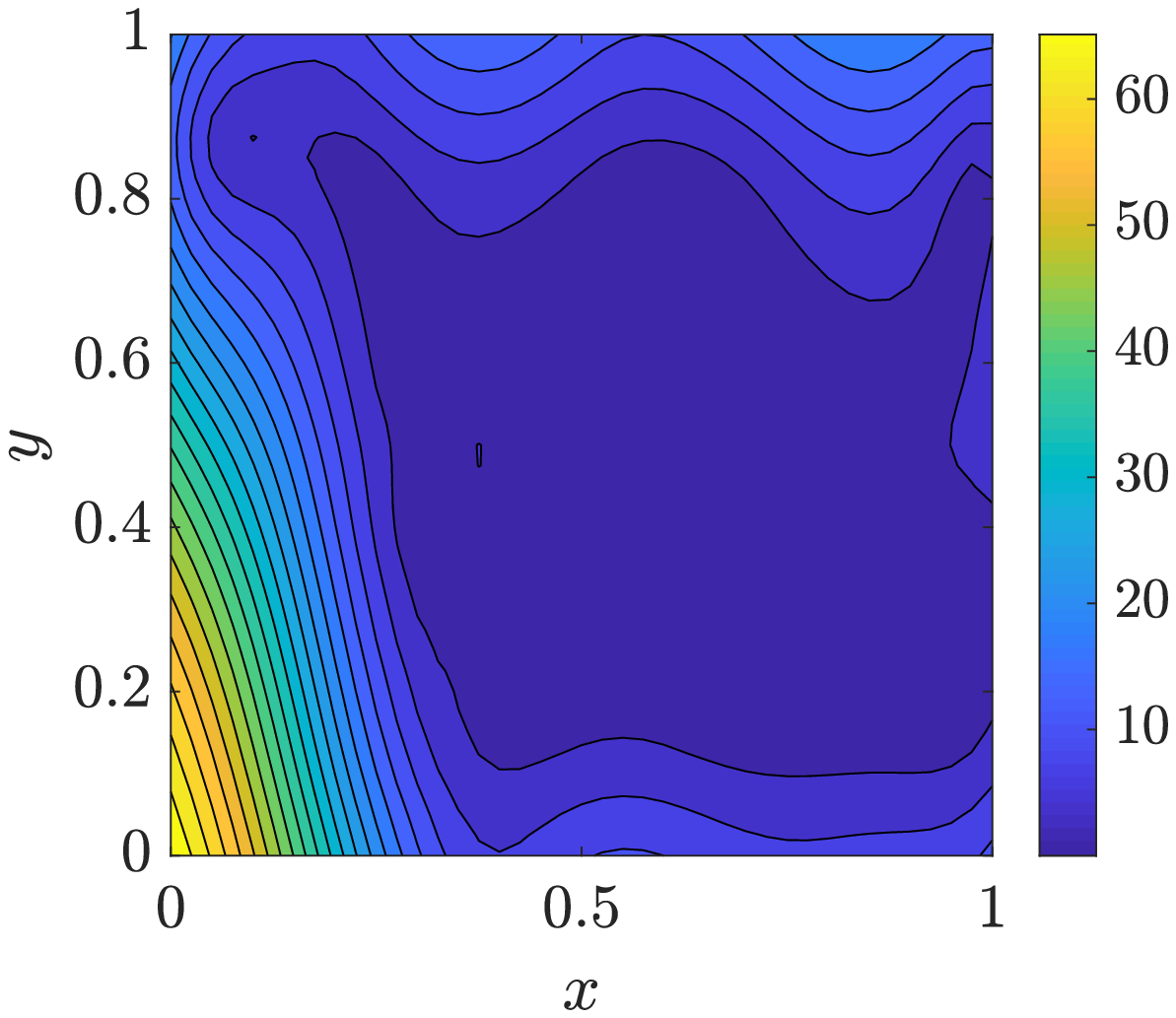}
        \caption{}
        \label{fig:2d_val_cokrig_std}
    \end{subfigure}
        \begin{subfigure}{0.32\textwidth}
        \includegraphics[width=\textwidth]{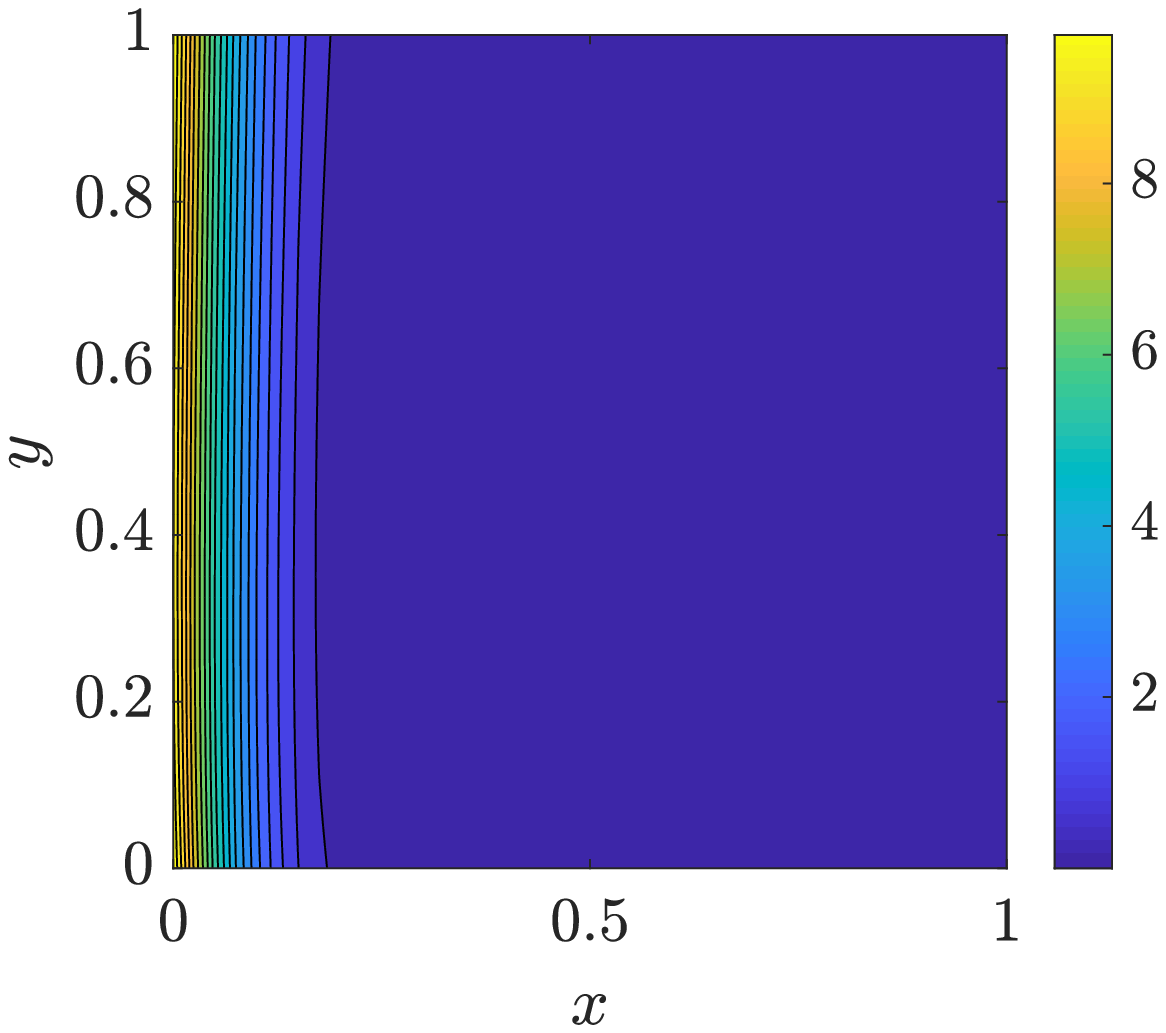}
        \caption{}
        \label{fig:2d_val_geck_std}
    \end{subfigure}
    \caption{The high-fidelity and low-fidelity function of the 2D problem and the  posterior prediction for the high-fidelity function. (a) The high-fidelity function, namely the modified Brainin function $f_H$ (contour) and observation locations (black squares). Posterior mean of QoI prediction by (b) Cokriging and (c) GE-Cokriging. (d) Low-fidelity function $f_L$ (contour) and observation locations (black dots). Posterior standard deviation of QoI by (e) Cokriging and (f) GE-Cokriging. Colored online.}
    \label{fig:2d_val}
\end{figure}

We extend the application of GE-Cokriging method in approximating a 2D function, namely a modified Branin function~\cite{forrester2008engineering}, given by
\begin{equation}
    f_H(x, y) = a(\Bar{x}_2 - b\Bar{x}_1^2 + c\Bar{x}_1 - r)^2 + g(1 - p)\cos(\Bar{x}_1) + g + qx,
\end{equation}
where 
\begin{equation*}
    \Bar{x}_1 = 15 x - 5, \Bar{x}_2 = 15 y, x \in [0, 1], y \in [0, 1],
\end{equation*}
with
\begin{equation*}
    a = 1, b = \frac{5.1}{4\pi ^2}, c = \frac{5}{\pi}, r = 6, g = 10, p = \frac{1}{8\pi},  q = 5,
\end{equation*}
and the low-fidelity function is constructed as follows,
\begin{equation}
    f_L(x, y) = Af_H(Bx + (1 - B), Cy),
\end{equation}
where A = 1.1, B = 0.95, C = 0.9. The contour of the modified Branin function $f_H$ that we aim to approximate is shown in Fig.~\ref{fig:2d_val_fe} and the contour for the low-fidelity function $f_L$ is shown in Fig.~\ref{fig:2d_val_fc}. The samples for high-fidelity observation locations $\mathbold{X}_H$ (black squares in Fig.~\ref{fig:2d_val_fe}) and low-fidelity observation locations $\mathbold{X}_L$ (black circles in Fig.~\ref{fig:2d_val_fc}) are randomly selected from the  uniformly spaced grid of size $41 \times 41$ on the domain $[0, 1] \times [0, 1]\in \mathbb{R}^2$. We note that $\mathbold{X}_H \subset \mathbold{X}_L$ as before.

We first compare the 
results of reconstructing $f_H$ by Cokriging and GE-Cokriging shown in Fig.~\ref{fig:2d_val}. It is clear that the posterior mean of GE-Cokriging (Fig.~\ref{fig:2d_val_geck}) is closer to $f_H$ than that of Cokriging (Fig.~\ref{fig:2d_val_cokrig}).
Also the degree of uncertainty is distinct as posterior standard deviation of Cokriging (Fig.~\ref{fig:2d_val_cokrig_std}) is one order of magnitude larger than that in GE-Cokriging (Fig.~\ref{fig:2d_val_geck_std}). 

\begin{figure}[!h]
    \centering
        \begin{subfigure}{0.32\textwidth}
        \includegraphics[width=\textwidth]{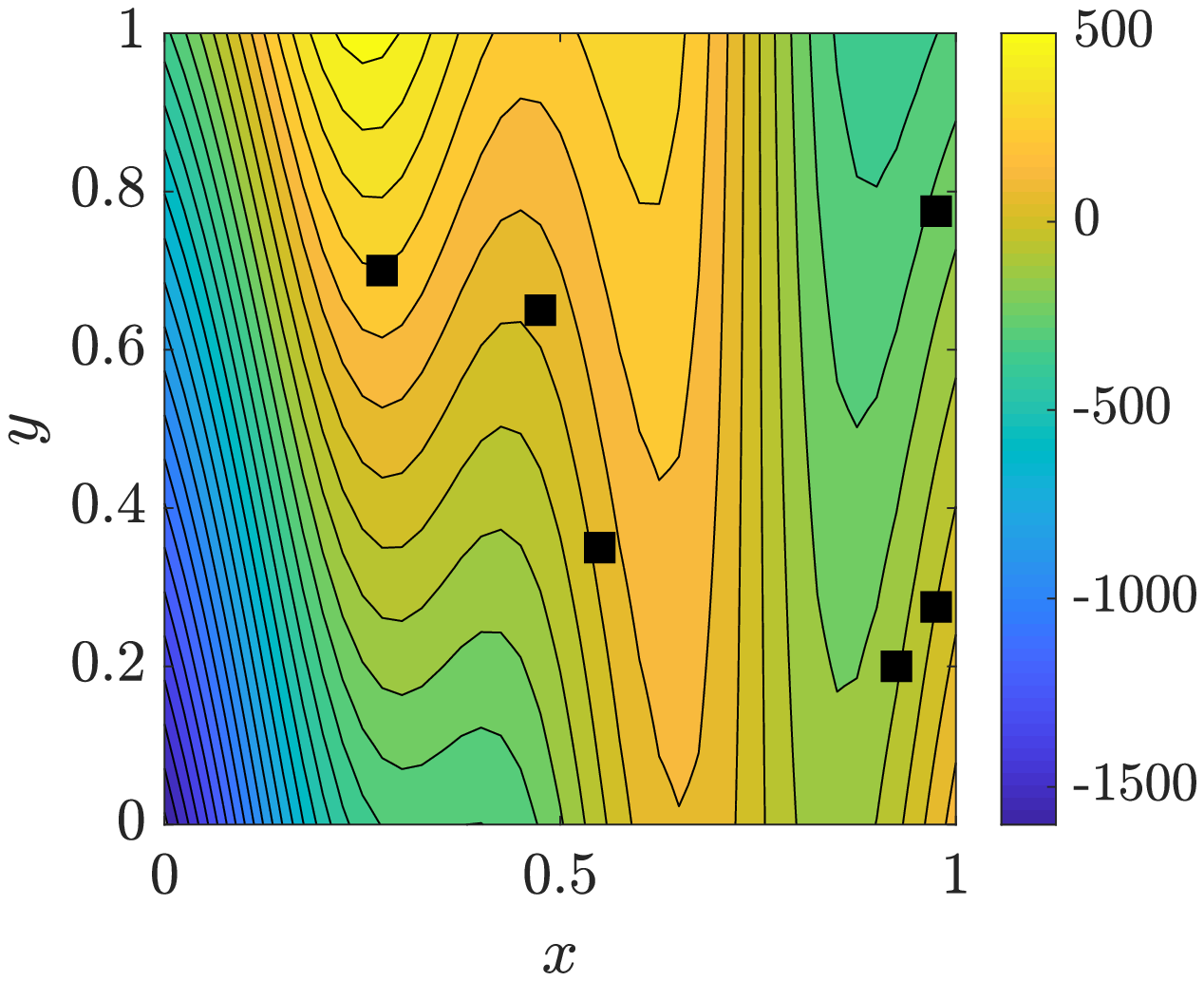}
        \caption{}
        \label{fig:2d_dfedx}
    \end{subfigure}
    \begin{subfigure}{0.32\textwidth}
        \includegraphics[width=\textwidth]{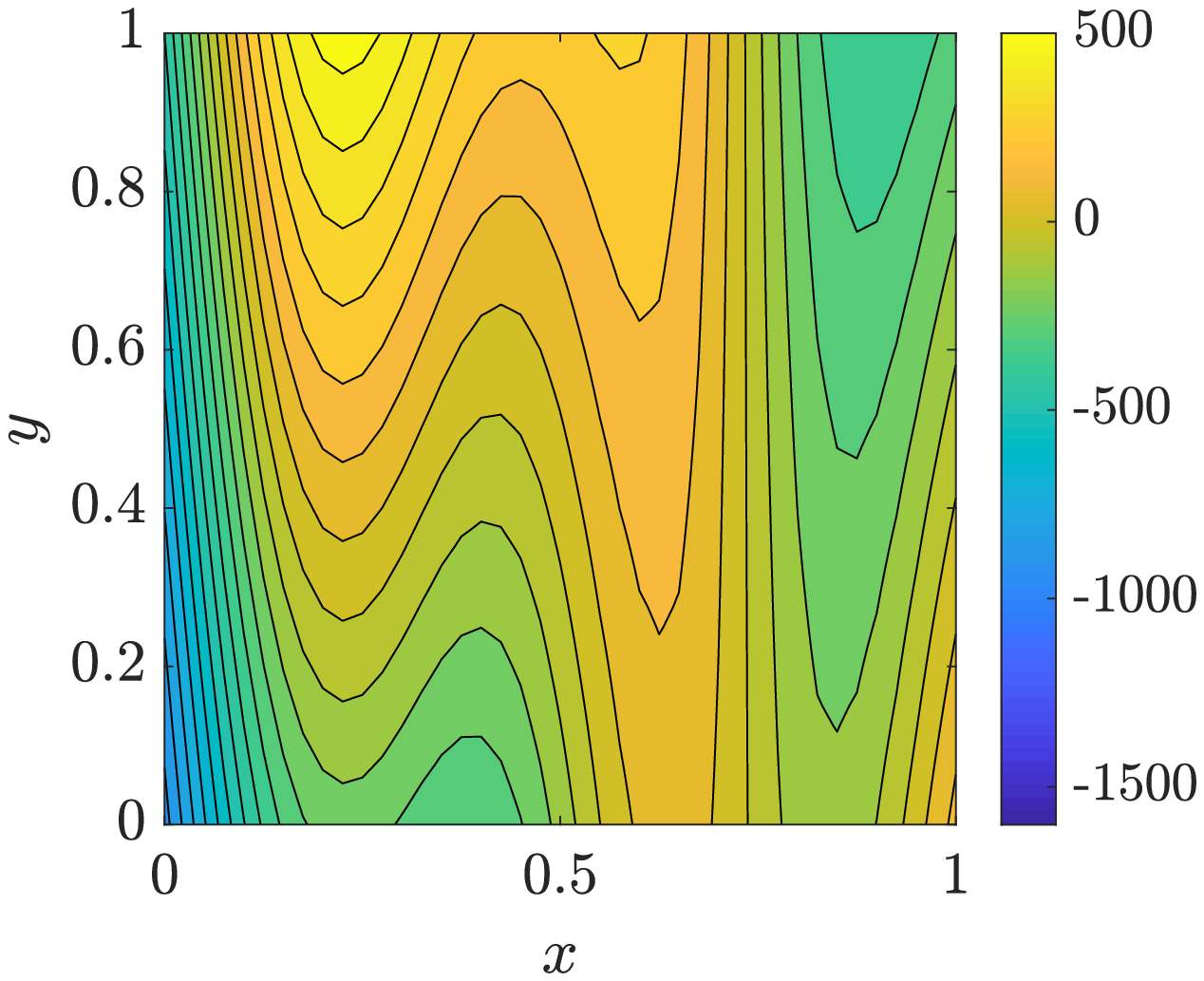}
        \caption{}
        \label{fig:2d_cokrig_dx}
    \end{subfigure}
        \begin{subfigure}{0.32\textwidth}
        \includegraphics[width=\textwidth]{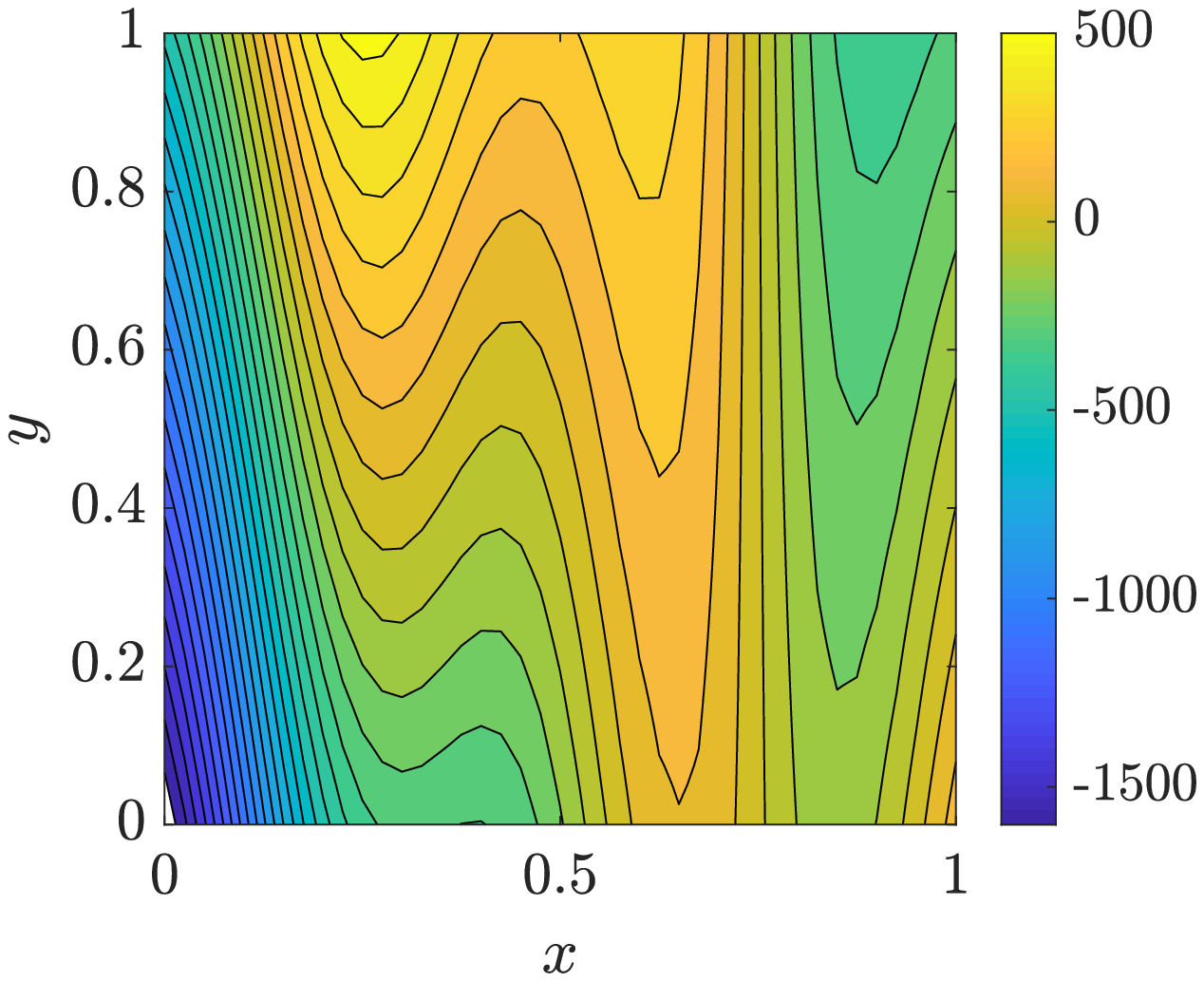}
        \caption{}
        \label{fig:2d_geck_dx}
    \end{subfigure}
    ~
    \begin{subfigure}{0.32\textwidth}
        \includegraphics[width=\textwidth]{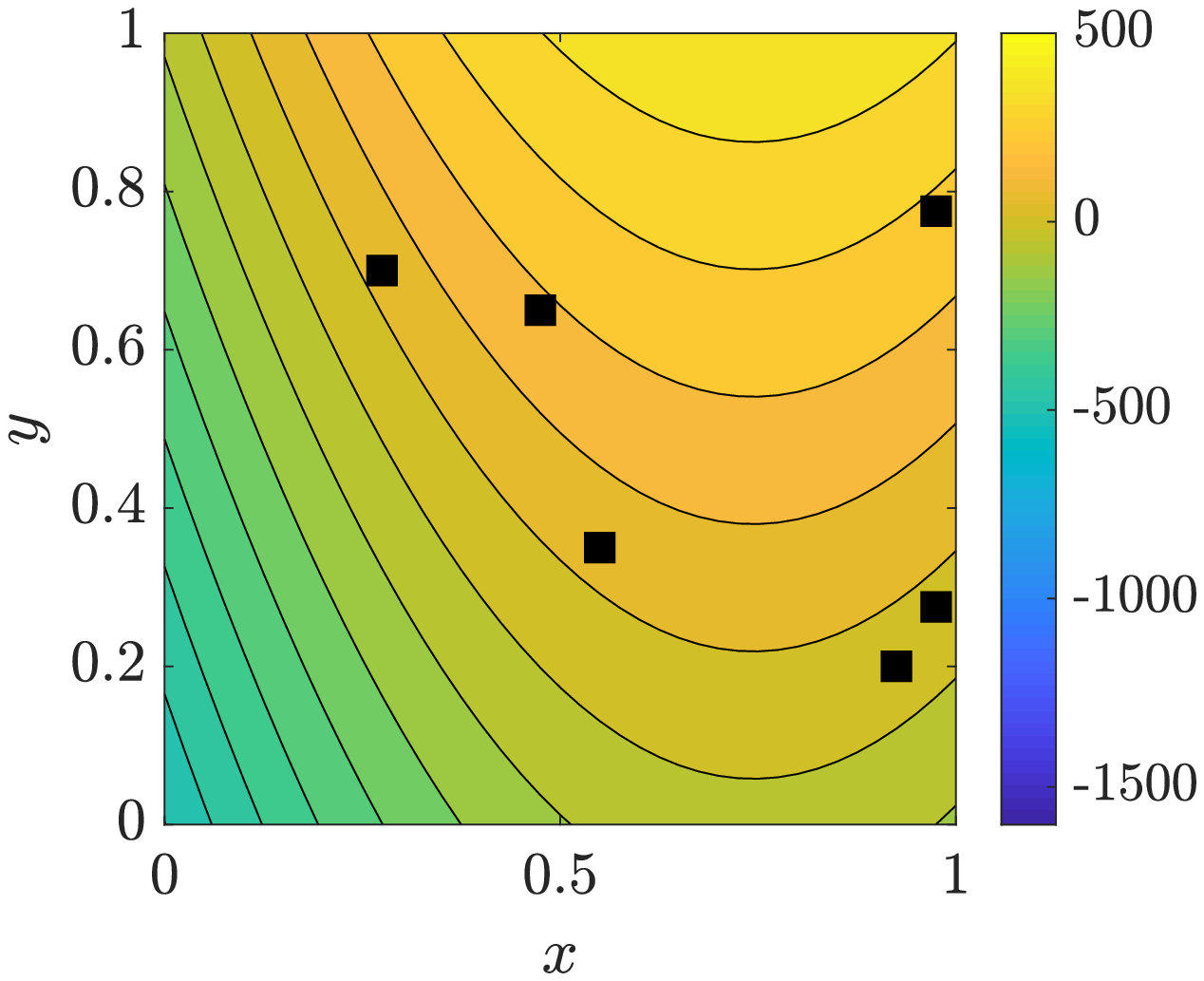}
        \caption{}
        \label{fig:2d_dfedy}
    \end{subfigure}
    \begin{subfigure}{0.32\textwidth}
        \includegraphics[width=\textwidth]{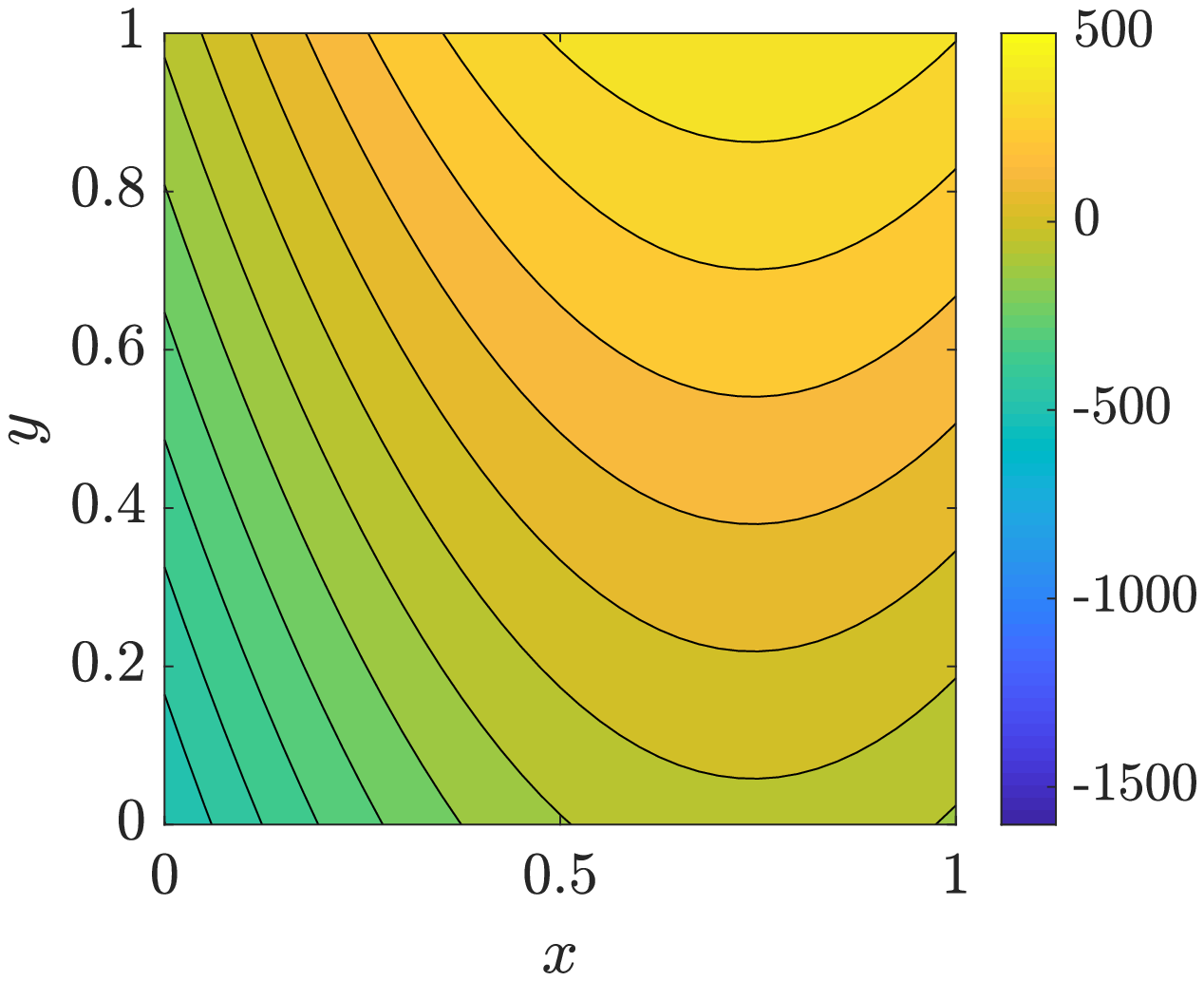}
        \caption{}
        \label{fig:2d_cokrig_dy}
    \end{subfigure}
        \begin{subfigure}{0.32\textwidth}
        \includegraphics[width=\textwidth]{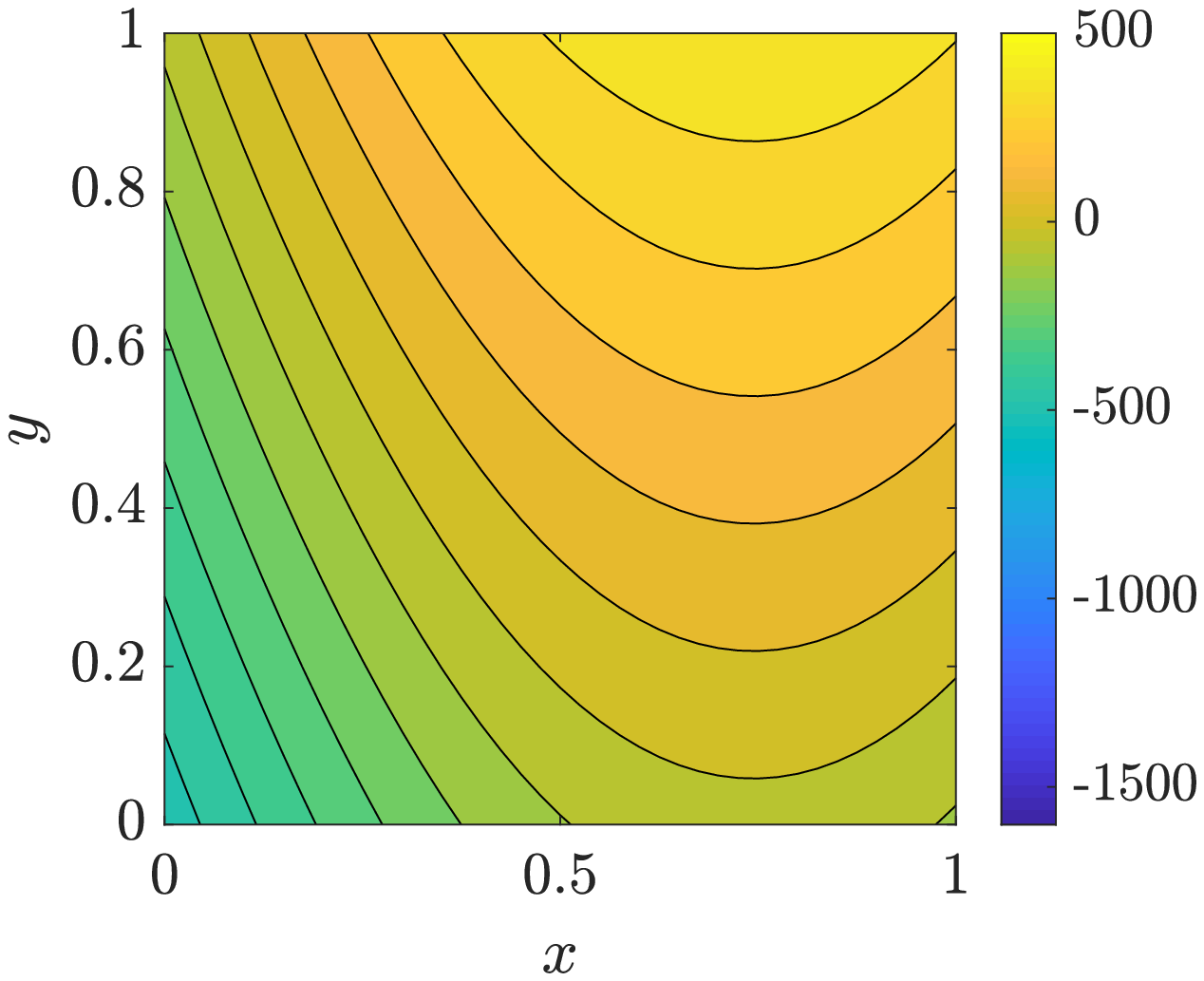}
        \caption{}
        \label{fig:2d_geck_dy}
    \end{subfigure}
    \caption{The high-fidelity gradients in $x$ and $y$ directions of the 2D problem and the corresponding posterior predictions. (a) The gradient of high-fidelity function in $x$ direction, $\frac{df_H}{dx}$ (contour) and high-fidelity samples (black squares) of gradient in $x$ direction. Posterior mean of gradient prediction in $x$ direction by (b) Cokriging and (c) GE-Cokriging. (d) The gradient of high-fidelity function in $y$ direction, $\frac{df_H}{dy}$ (contour) and high-fidelity samples (black squares) of gradient in $y$ direction. Posterior mean of gradient prediction in $y$ direction by (e) Cokriging and (f) GE-Cokriging. Colored online.}
    \label{fig:2d_grad}
\end{figure}

Next, we compare the prediction of gradients by Cokriging and GE-Cokriging. Fig.~\ref{fig:2d_dfedx} and Fig.~\ref{fig:2d_dfedy} profile contours of exact $\frac{\partial f_H}{\partial x}$ and $\frac{\partial f_H}{\partial y}$, respectively. For predicting $\frac{\partial f_H}{\partial x}$, GE-Cokriging (Fig.~\ref{fig:2d_geck_dx}) shows higher accuracy globally while Cokriging (Fig.~\ref{fig:2d_cokrig_dx}) can not result in accurate prediction in the lower left corner, where the available observation data is rare. As for $\frac{\partial f_H}{\partial y}$, since the target function is relatively smooth, both Cokriging (Fig.~\ref{fig:2d_cokrig_dy}) and GE-Cokriging (Fig.~\ref{fig:2d_geck_dy}), are capable of obtaining accurate prediction, while GE-Cokriging still outperforms Cokriging in the sense of the total RMSE recorded in Tab.~\ref{tab:error_table}. 

\subsection{Underdamped oscillator}
We consider a driven harmonic oscillator described by the following second order ODE:
\begin{equation} \label{eq:osci0}
  \begin{cases}
  m\ddot{x} + c\dot{x} + kx = F(t), \\
  x(0) = 1, \quad \dot{x}(0)=0,
  \end{cases}
\end{equation}
where $m$ is the mass, $c$ is the damping coefficient, $k$ is a constant (e.g.,
elasticity coefficient of a string), and $F(t)$ is the external force. We rewrite
the ODE in Eq.~\eqref{eq:osci0} as
\begin{equation}
  \label{eq:osci1}
  \ddot{x} + 2\zeta\omega_0\dot{x} + \omega_0^2x = \dfrac{F(t)}{m},
\end{equation}
where $\omega_0=\sqrt{\dfrac{k}{m}}$ is the undamped angular frequency, and
$\zeta=\dfrac{c}{2\sqrt{mk}}$ is the damping ratio. We set $\zeta=1/\sqrt{37}$
and $\omega_0=\dfrac{6}{\sqrt{1-\zeta^2}}$ in this study. The external force is
set as the step response:
\begin{equation}\label{eq:force}
  \dfrac{F(t)}{m} = 
  \begin{cases}
    \omega_0^2, & t\geq 0, \\
    0,  & t < 0.
  \end{cases}
\end{equation}
The analytical solution to Eq.~\eqref{eq:osci0} is 
\begin{equation}
  x_H(t) = \me^{-\zeta\omega_0 t}\dfrac{\sin(\sqrt{1-\zeta^2}\omega_0 t
  +\varphi)}{\sin \varphi},\quad \varphi = \arccos\zeta, 
\end{equation}
and the velocity is
\begin{equation}
  \dot{x}_H(t) = -\dfrac{\omega_0\me^{-\zeta\omega_0 t}}{\sin\varphi}
      \left[\zeta\sin(\sqrt{1-\zeta^2}\omega_0 t +\varphi) 
          - \sqrt{1-\zeta^2}\cos(\sqrt{1-\zeta^2}\omega_0 t +\varphi)\right]. 
\end{equation}
The low-fidelity model is a simple harmonic oscillator model:
\begin{equation} \label{eq:osci2}
  \begin{cases}
  m\ddot{x} + kx = 0, \\
  x(0) = 1, \quad \dot{x}(0)=0,
  \end{cases}
\end{equation}
which is equivalent to setting $\zeta=0$ and $F(t)=0$ in Eq.~\eqref{eq:osci1}.
The analytical solution to the low-fidelity model is
\begin{equation}
  x_L(t) = \cos(\omega_0 t), 
\end{equation}
and the velocity is
\begin{equation}
  \dot{x}_L(t) = -\omega_0\sin(\omega_0 t). 
\end{equation}

\begin{figure}[!ht]
    \centering
    \begin{subfigure}{0.45\textwidth}
    \centering
        \includegraphics[width=\textwidth]{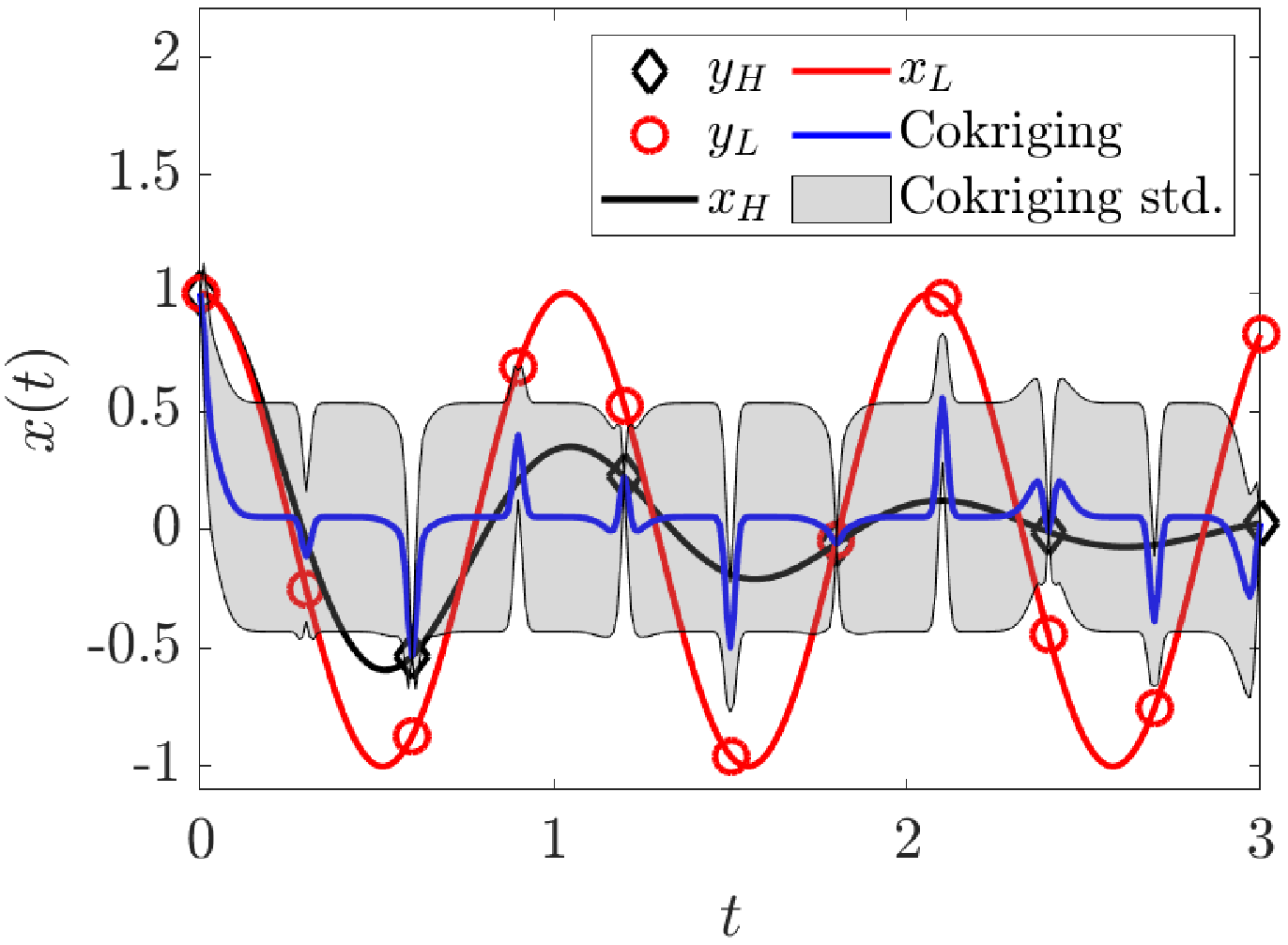}
        \caption{}
        \label{fig:damper_val_cokrig}
    \end{subfigure}
    \hfill
    \begin{subfigure}{0.45\textwidth}
    \centering
        \includegraphics[width=\textwidth]{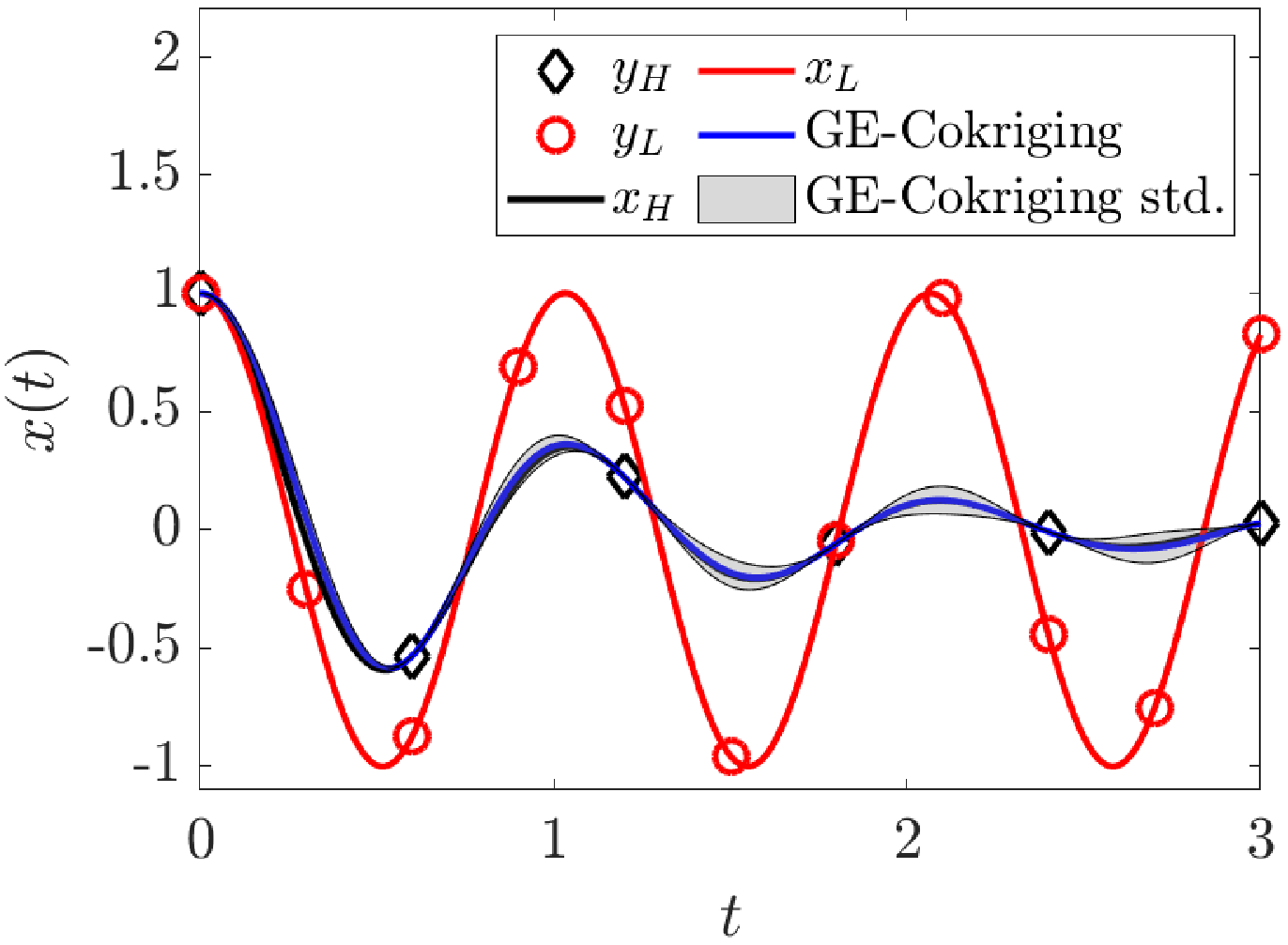}
        \caption{}
        \label{fig:damper_val_geck}
    \end{subfigure}
    ~
    \begin{subfigure}{0.45\textwidth}
    \centering
        \includegraphics[width=\textwidth]{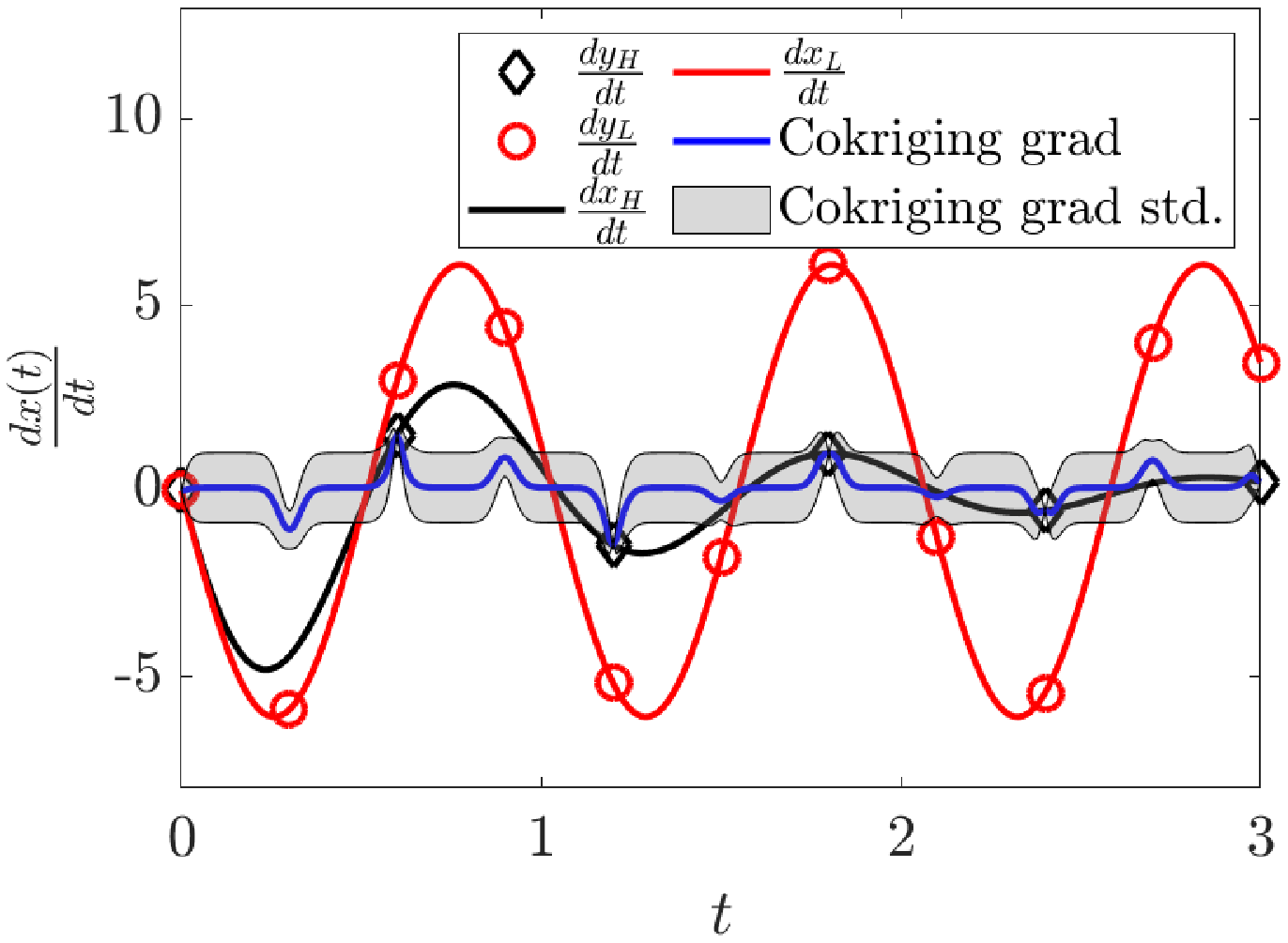}
        \caption{}
        \label{fig:damper_grad_cokrig}
    \end{subfigure}
    \hfill
    \begin{subfigure}{0.45\textwidth}
    \centering
        \includegraphics[width=\textwidth]{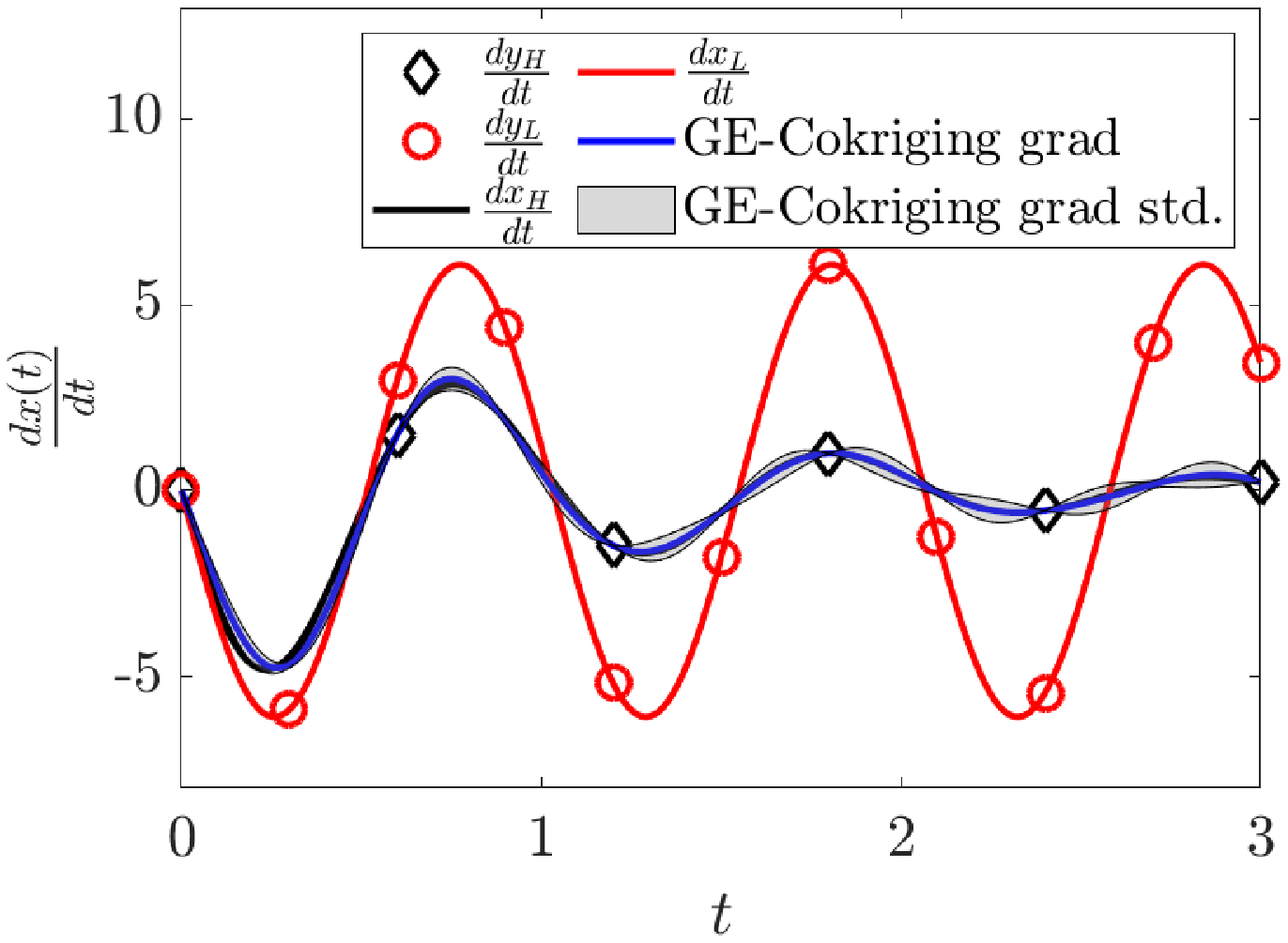}
        \caption{}
        \label{fig:damper_grad_geck}
    \end{subfigure}
    ~
    \begin{subfigure}{0.45\textwidth}
    \centering
        \includegraphics[width=\textwidth]{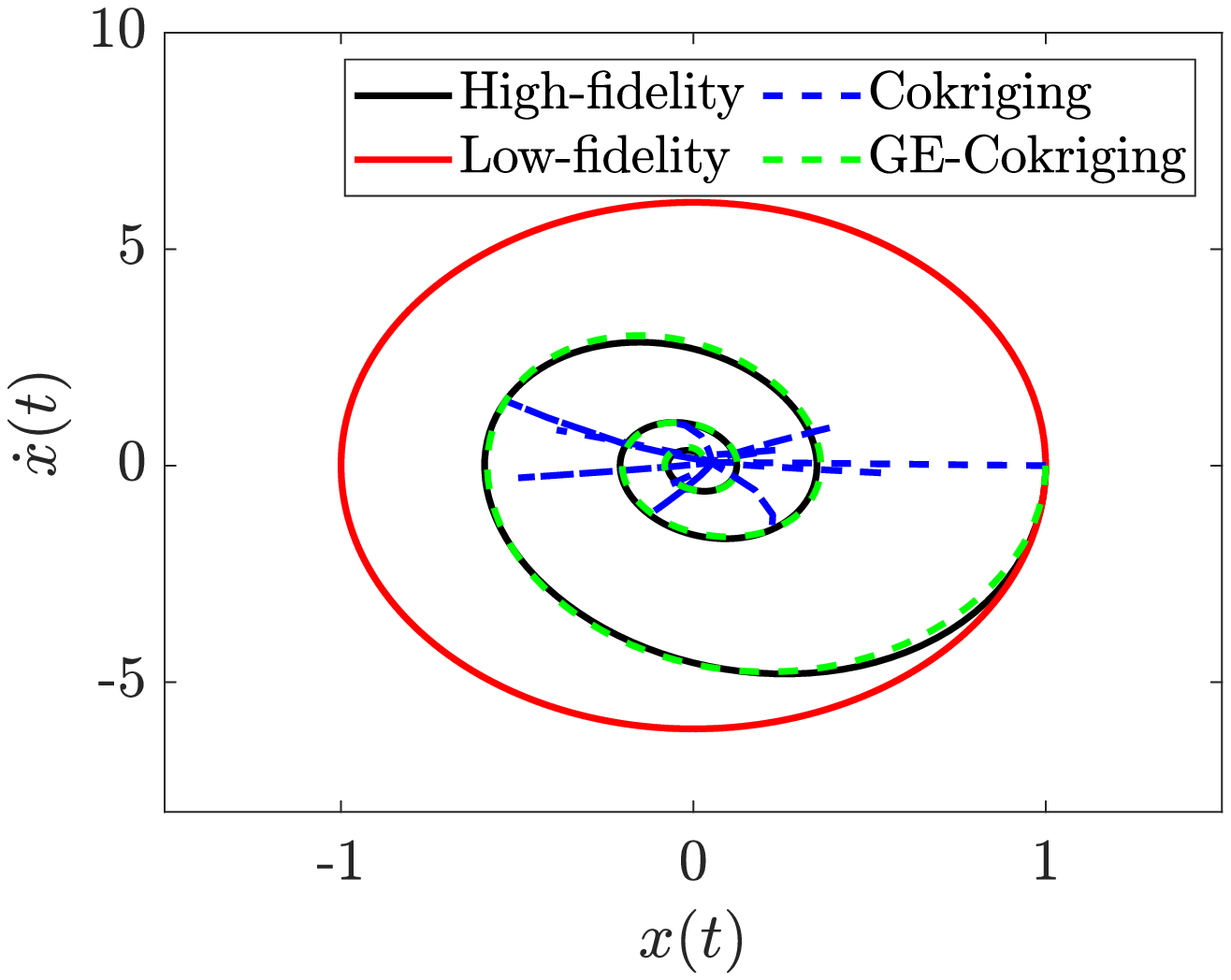}
        \caption{}
        \label{fig:damper_phase_diagram}
    \end{subfigure}
    \caption{Prediction of the trajectory (QoI), velocity (gradient of QoI) and the phase diagram of an underdamped oscillator. Prediction of the posterior mean (blue solid lines) and standard deviation (grey shaded area) of the trajectory $x_H(t)$ by (a) Cokriging and (b) GE-Cokriging. Prediction of the posterior mean (blue solid lines) and standard deviation (grey shaded area) of the velocity $\frac{dx_H(t)}{dt}$ by (c) Cokriging and (d) GE-Cokriging. (e) Prediction of phase diagram by Cokriging (blue dashed line) and that by GE-Cokriging (black dashed line). Black diamonds denote high-fidelity observations, red circles denote low-fidelity observations, black solid lines denote the high-fidelity models and red solid lines denote low-fidelity models. Colored online.}
    \label{fig:damper}
\end{figure}

The observation locations for high- and low-fidelity models are set as $T_H=\{0.6j\}_{j=0}^5$ and $T_L=\{0.3j\}_{j=0}^{10}$, respectively.
We compare the constructed trajectory $x(t)$ and velocity $\dot{x}(t)$ on $[0,3]$ by Cokriging and GE-Cokriging in Fig.~\ref{fig:damper}. Cokriging again shows worse performance both for prediction of QoI (Fig.~\ref{fig:damper_val_cokrig}) and gradient (Fig.~\ref{fig:damper_grad_cokrig}) marked by significant deviations from the true values as well as large uncertainties at locations distant from observation locations, while GE-Cokriging manages to reconstruct the trajectory (Fig.~\ref{fig:damper_val_geck}) and velocity (Fig.~\ref{fig:damper_grad_geck}) of the oscillator well with small standard deviations. The overlapping between trajectory-velocity phase diagram by GE-Cokriging and the exact phase diagram (Fig.~\ref{fig:damper_phase_diagram}) emphasizes that GE-Cokriging can provide accurate predictions for QoI and the corresponding gradients simultaneously, while Cokriging failed to. We also note that Cokring suffers from singularity of the covariance matrix again, while GE-Cokriging doesn't have this concern.

\subsection{Sensitivity of a power grid system}
We now consider the relationship between the power input of a generator bus, denoted as $x$, and real-time power factor of a load bus, as $f(x)$, in a large-scale power system from IEEE 118 bus test case~\cite{christie1993}. We use MATPOWER~\cite{zimmerman2011matpower}, which provides a model for the IEEE 118 bus test case, to run simulations and generate sample points. The $f_H(x)$ and $f_L(x)$ represent the alternative current (AC) and direct current (DC) models approximating $f(x)$, respectively.

The observation locations for Cokriging and GE-Cokriging
consist of $51$ low-fidelity samples from DC model on 
$X_L = \{20 + 2j\}_{j=0}^{50}$ and five samples from AC model on $X_H = \{40, 48, 72, 98, 116\}$ (again, $X_H \subset X_L$). In addition to reconstructing $f_H$ accurately, 
estimating the change of power factor of a load bus in response to the change of power input of a generate bus, i.e., the sensitivity of $f$ with respect to $x$, is important for safety or energy-efficiency consideration. This change is reflected by the derivative of $f(x)$, i.e., $\frac{\dif f(x)}{\dif x}$. Therefore, we aim to approximate both $f_H$ and its derivative.
Here we use finite-difference method to obtain $\frac{\dif f_H}{\dif x}$ and $\frac{\dif f_L}{\dif x}$ at $X_H$ and $X_L$, respectively, and the step size is $0.25$.

\begin{figure}[!ht]
    \centering
    \begin{subfigure}{0.45\textwidth}
    \centering
        \includegraphics[width=\textwidth]{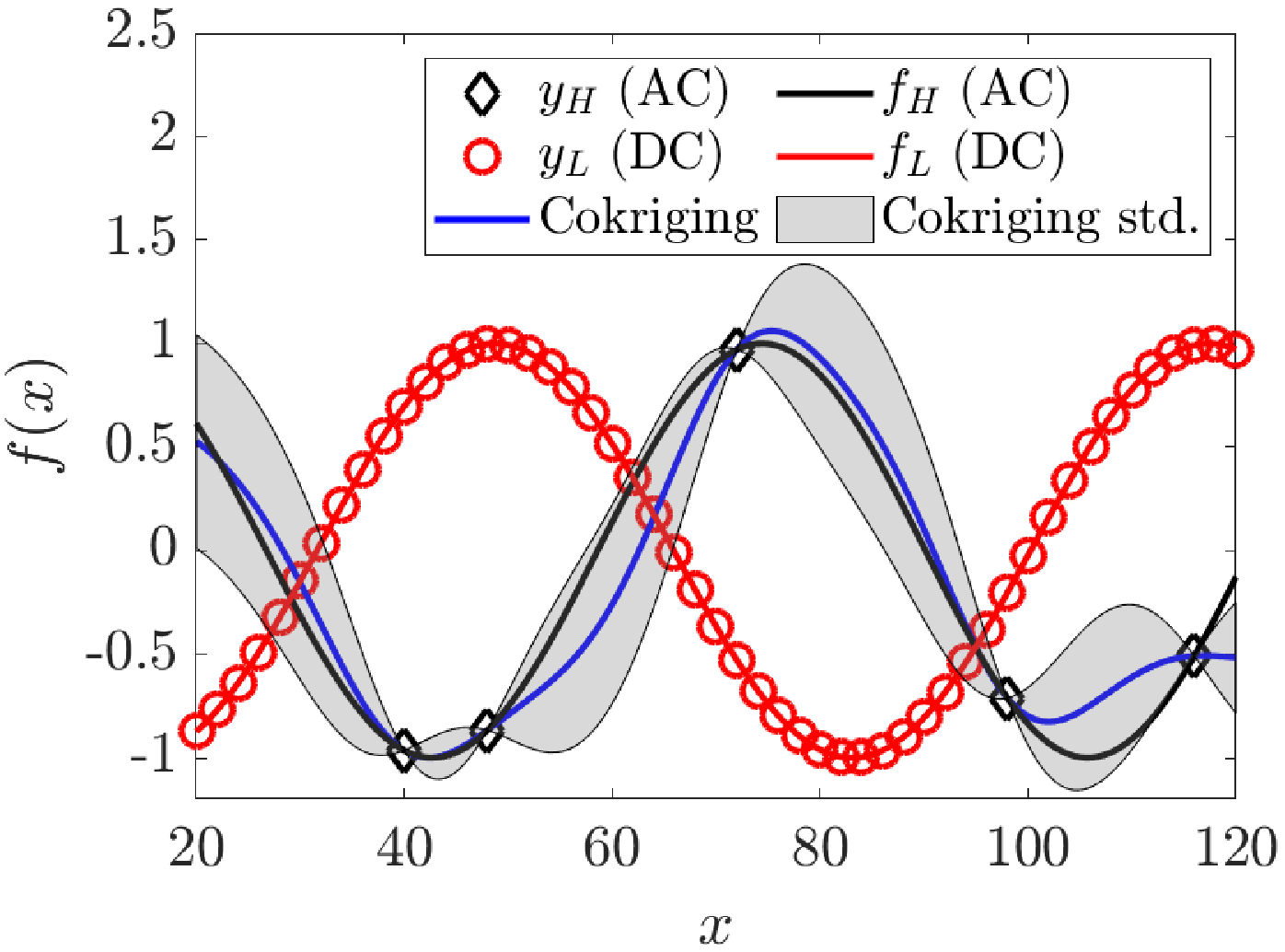}
        \caption{}
        \label{fig:power_val_cokrig}
    \end{subfigure}
    \hfill
    \begin{subfigure}{0.45\textwidth}
    \centering
        \includegraphics[width=\textwidth]{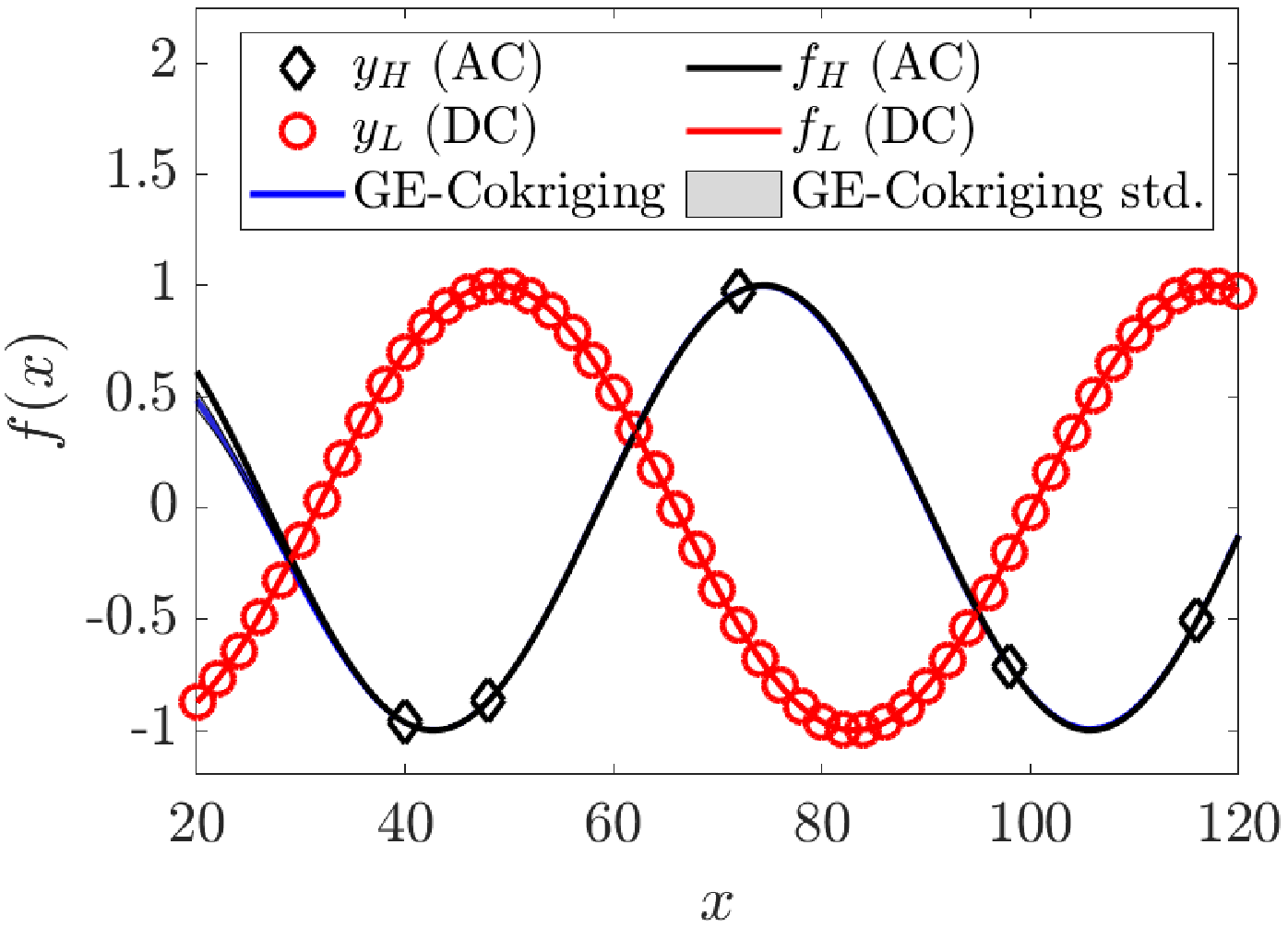}
        \caption{}
        \label{fig:power_val_geck}
    \end{subfigure}
    \begin{subfigure}{0.45\textwidth}
    \centering
        \includegraphics[width=\textwidth]{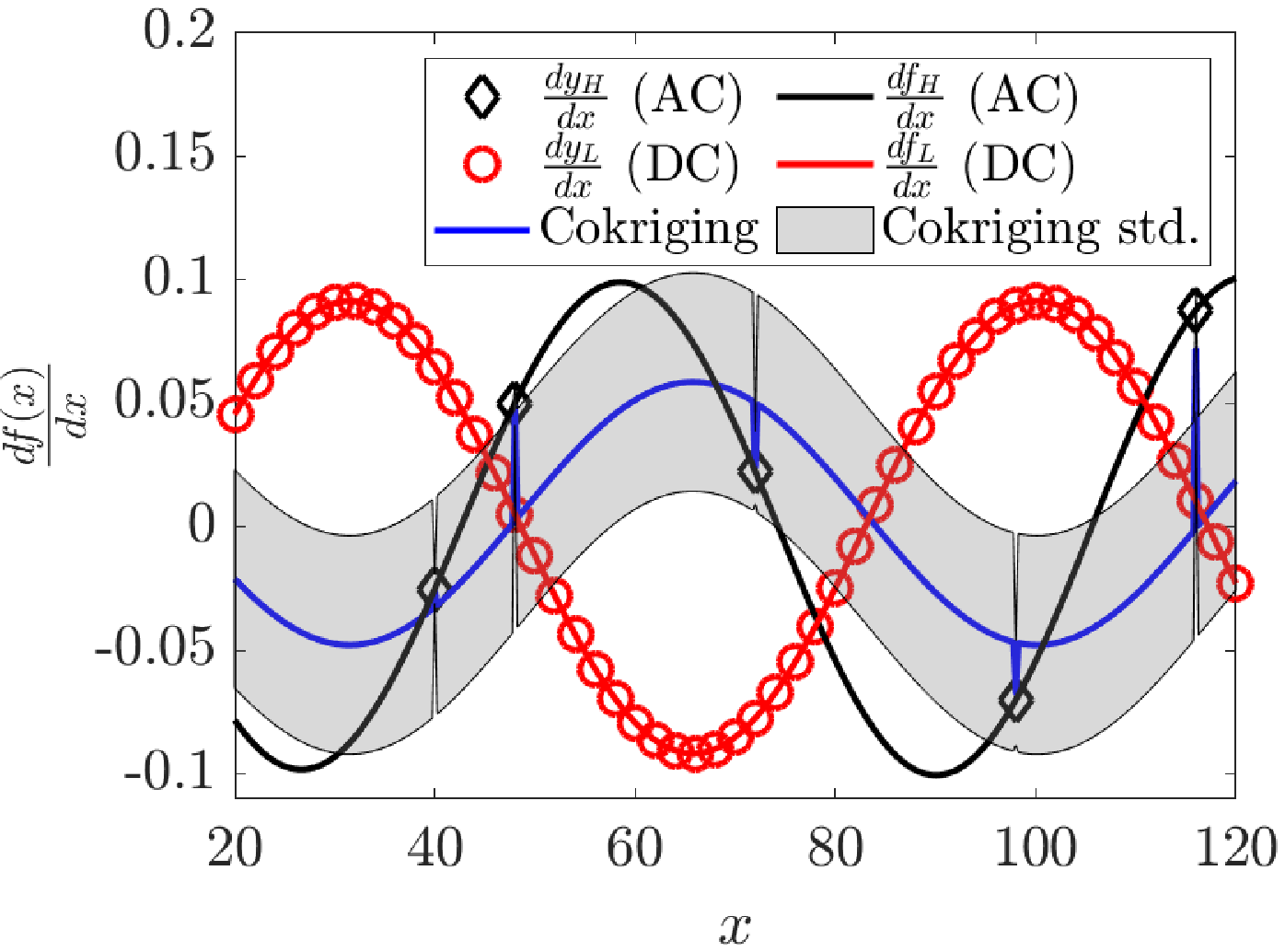}
        \caption{}
        \label{fig:power_grad_cokrig}
    \end{subfigure}
    \hfill
    \begin{subfigure}{0.45\textwidth}
    \centering
        \includegraphics[width=\textwidth]{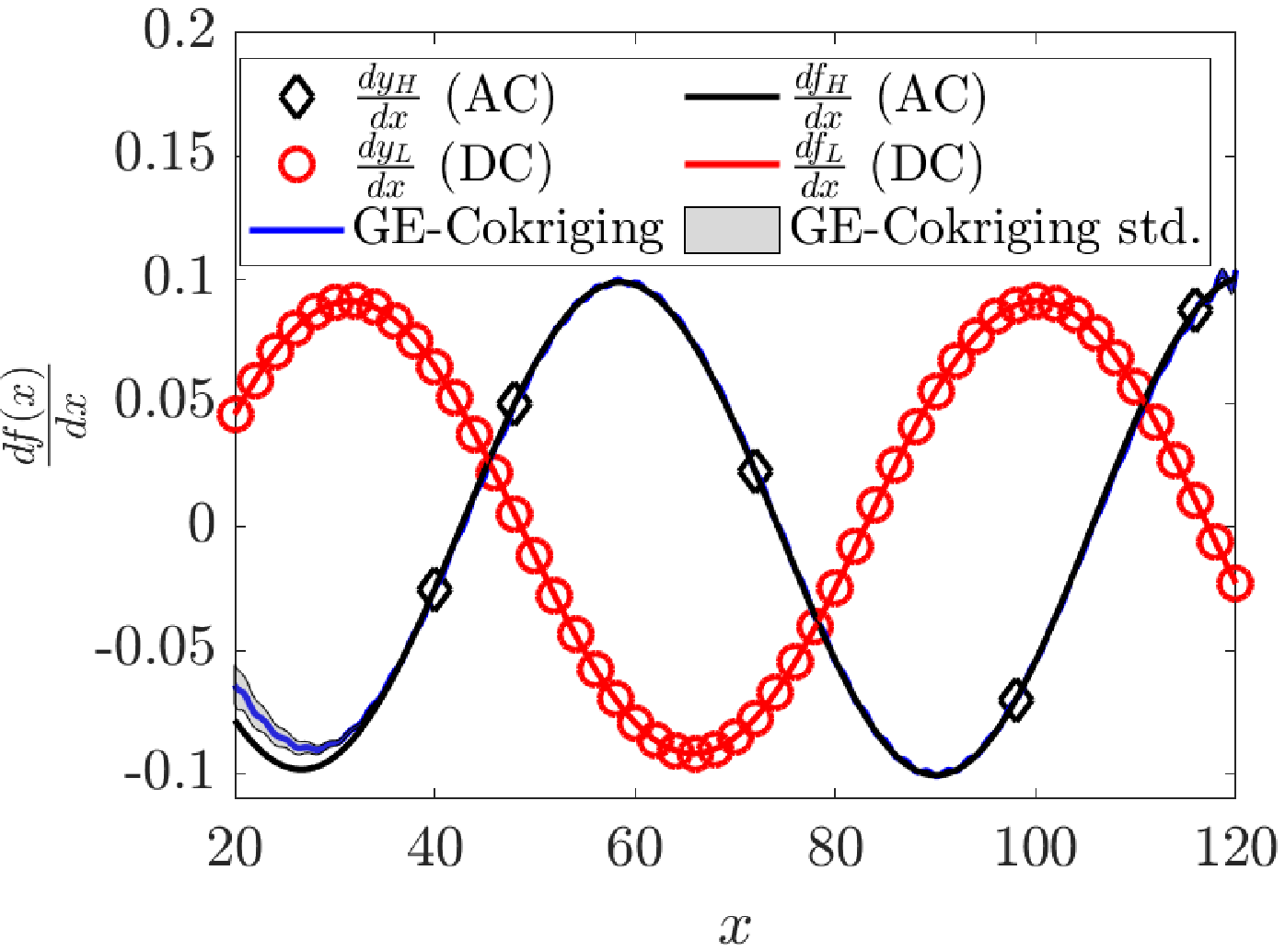}
        \caption{}
        \label{fig:power_grad_geck}
    \end{subfigure}
    \caption{Prediction of the relationship between the power input of a generator bus $x$ and real-time power factor of a load bus $f_H(x)$ by an AC model. Prediction of the posterior mean (blue solid lines) and standard deviation (grey shaded area) of $f_H(x)$ by (a) Cokriging and (b) GE-Cokriging. Prediction of the posterior mean (blue solid lines) and standard deviation (grey shaded area) of gradient of QoI $\frac{df_H(x)}{dx}$ by (c) Cokriging and (d) GE-Cokriging. Black diamonds denote high-fidelity observations, red circles denote low-fidelity observations, black solid lines denote the high-fidelity models and red solid lines denote the low-fidelity models. Colored online.}
    \label{fig:power}
\end{figure}

The results in Fig.~\ref{fig:power} suggest that the Cokriging method can approximate $f_H$
with noticeable standard deviations (Fig.~\ref{fig:power_val_cokrig}), but it fails to reconstruct $\frac{\dif f_H}{\dif x}$ 
(Fig.~\ref{fig:power_grad_cokrig}). On the other hand, GE-Cokriging can reconstruct both $f_H$ (Fig.~\ref{fig:power_val_geck}) and $\frac{\dif f_H}{\dif x}$ (Fig.~\ref{fig:power_grad_geck}) accurately with rather small uncertainty, and the only noticeable discrepancy appears near the left boundary because that region is far from available data.
Unlike other cases, here we notice the occurrence of wiggling in the high-fidelity gradient prediction by GE-Cokriging. 
This is caused by the aliasing error as we used finite-different method to approximate the gradient functions, recall that no wiggling is observed in previous examples where the gradients are observed directly. Again, reconstructing the gradient using Cokriging suffers from the sigularity of the covariance matrix as shown in Fig.~\ref{fig:power_grad_cokrig}, whereas GE-Cokriging doesn't have this concern (see Fig.~\ref{fig:power_grad_geck}).

\subsection{Quantitative comparison and computational efficiency}
\label{subsec:err}
To analyze and compare the accuracy and efficiency among Cokriging, GE-Kriging and GE-Cokriging, we run simulations for five times with random initial conditions for each numerical example, and list the relative mean squared errors for QoI prediction and gradient of QoI prediction in Tab.~\ref{tab:error_table}. The numerical simulations were performed on the same laptop with Intel(R) Core(TM) i7-8550U CPU @ 1.80GHz. We recorded the time for each separate run and computed the corresponding mean and standard deviation from these 5 runs for each example (see Tab.~\ref{tab:time_table}). 

The results in Tab.~\ref{tab:error_table} show that GE-Cokriging outperforms Cokriging and GE-Kriging in terms of relative mean squared error for all examples presented. We note that in GE-Kriging, only high-fidelity QoI data (including high-fidelity gradient data) was used for training. GE-Cokriging improves accuracy in all cases compared to Cokriging, which is consistent to the visual observations shown in each numerical example. It is also worth noting that the relative mean squared errors by Cokriging are almost one order of magnitude higher than those by GE-Cokriging in most of the cases. The errors in the prediction of QoIs by GE-Kriging are several times larger than those by GE-Cokriging, and the prediction of gradients by GE-Kriging are even worse than those by GE-Cokriging, in all examples. Hence, among these three methods compared, GE-Cokriging is able to maintain a robust prediction result both in terms of QoI and in terms of the gradient of QoI simultaneously, while the other two methods can not obtain comparable results.
This further verifies that the information of QoI and its gradients can be strongly correlated, and hence is of great help to improve the accuracy of GPR methods when used jointly. 

\begin{table}[h!]
    \centering
    \scriptsize
    \resizebox{\textwidth}{!}{\begin{tabular}{ccccccc}
        \toprule
         Case & Cokriging & GE-Kriging & GE-Cokriging & Cokriging ($\nabla$) & GE-Kriging ($\nabla$) & GE-Cokriging ($\nabla$)\\
         \midrule
         1D1     & $0.1146\pm 1.49$e-2
                 & $0.7534\pm 5.17$e-6
                 & $0.0138\pm 9.98$e-5 
                 & $0.9565\pm 1.84$e-2
                 & $0.5985\pm 3.85$e-6
                 & $0.0221\pm 1.49$e-4\\
         1D2     & $0.5325\pm 1.47$e-5 
                 & $0.7534\pm 6.30$e-6
                 & $0.1254\pm 4.18$e-5 
                 & $0.8964\pm 2.97$e-5 
                 & $0.5986\pm 4.83$e-6
                 & $0.0973\pm 3.13$e-6\\
         2D$^*$  & $0.3152\pm 2.21$e-1  
                 & $0.2471\pm 1.60$e-1
                 & $0.0292\pm 8.86$e-3 
                 & $0.3101\pm 1.88$e-1
                 & $0.4062\pm 2.03$e-1
                 & $0.0798\pm 3.26$e-2\\
         2D$^{**}$& -  
                  & -
                  & - 
                  & $0.1341\pm 1.78$e-2
                  & $0.3342\pm 1.39$e-1
                  & $0.0114\pm 5.81$e-3\\
         Oscillator & $0.9224\pm 2.15$e-2
                    & $0.1259\pm 3.27$e-6
                    & $0.0926\pm 2.40$e-5 
                    & $1.0771\pm 3.92$e-2
                    & $0.2639\pm 1.78$e-1
                    & $0.0993\pm 2.49$e-5\\
         Power      & $0.2451\pm 4.79$e-4  
                    & $0.1888\pm 1.57$e-5
                    & $0.0363\pm 4.87$e-6
                    & $0.7391\pm 1.08$e-2
                    & $0.2413\pm 2.82$e-5
                    & $0.0522\pm 7.23$e-6\\
    \bottomrule
    \end{tabular}}
    \caption{Relative mean squared error (mean$\pm$standard deviation) of QoI and the corresponding gradients for each numerical example averaged over 5 separate runs with random parameters initialization by Cokriging, GE-Kriging and GE-Cokriging. $*$ denotes gradient in $x$ direction and $**$ denotes gradient in $y$ direction. $\nabla$ denotes prediction of the gradient of QoI.}
    \label{tab:error_table}
\end{table}

Tab.~\ref{tab:time_table} shows that GE-Kriging and GE-Cokriging are more time-efficient compared to Cokriging, which is suggested by the fact that the prediction of gradients with GE-Kriging and GE-Cokriging take a rather small amount of time compared to Cokriging method. This is due to the fact that GE-Kriging and GE-Cokriging integrate both QoI data and the corresponding gradient data in the training step and hence provides prediction of QoI as well as the gradient on the new locations simultaneously in the predicting step. Whereas, Cokriging requires construction of a model for gradient data separately. Hence, the time for the prediction of the gradients by GE-Kriging and GE-Cokriging, i.e., the last two columns in Tab.~\ref{tab:time_table}, are for prediction only and is relatively short. It is also noticed that the time consumption of GE-Kriging is smaller than that of GE-Cokriging, recall that GE-Kriging only used high-fidelity information while GE-Cokriging used both high-fidelity and low-fidelity information, which lead to a larger covariance matrix in GE-Cokriging compared to that in GE-Kriging. Although GE-Cokriging generally requires longer time in the training step, almost doubles Cokriging's training time, the total time cost of GE-Cokriging in QoI and gradients prediction is almost the same as that of Cokriging. Considering the significant improvement in accuracy and robustness, we can conclude that GE-Cokriging is an accurate and efficient approach to obtain prediction both QoI and its gradients simultaneously.
\begin{table}[h!]
    \centering
    \scriptsize
    \resizebox{\textwidth}{!}{\begin{tabular}{ccccccc}
    \toprule
         Case ID & Cokriging  & GE-Kriging & GE-Cokriging & Cokriging ($\nabla$) & GE-Kriging ($\nabla$) & GE-Cokriging ($\nabla$)\\
         \midrule
         1D1  & $1.5702\pm 1.33$e-2
              & $0.4193\pm 4.196$e-2
              & $2.0945\pm 1.28$e-1
              & $1.2128\pm 1.42$e-1
              & $0.0042\pm 4.42$e-4
              & $0.0156\pm 1.56$e-3\\
         1D2  & $0.8337\pm 1.06$e-1  
              & $0.4452\pm 3.06$e-2
              & $1.0767\pm 8.82$e-2
              & $0.8329\pm 9.54$e-2 
              & $0.0036\pm 6.76$e-4
              & $0.0142\pm 1.18$e-3\\
         2D$^*$   & $2.0945\pm 7.08$e-1
                  & $0.7623\pm 5.26$e-2
                  & $3.5935\pm 7.42$e-1
                  & $1.3417\pm 2.96$e-1
                  & $0.1245\pm 2.93$e-2
                  & $0.7502\pm 1.72$e-2 \\
         2D$^{**}$  & - & - & -
                    & $1.6846\pm 3.71$e-2  & - & -\\
         Oscillator & $0.6402\pm 3.75$e-2  
                    & $0.3321\pm 1.84$e-1
                    & $1.1046\pm 9.08$e-2
                    & $0.6926\pm 3.34$e-2
                    & $0.0095\pm 8.79$e-4
                    & $0.0074\pm 1.96$e-3\\
          Power & $0.4127\pm 1.15$e-2 
               & $0.4933\pm 8.01$e-2
               & $2.4326\pm 1.34$e-1
               & $0.9492\pm 3.21$e-3
               & $0.0119\pm 5.62$e-3
               & $0.0198\pm 2.56$e-3\\ 
    \bottomrule
    \end{tabular}}
    \caption{Runtime (mean$\pm$standard deviation) of predicting QoI and its gradients for each numerical example averaged over 5 separate runs with random parameters initialization by Cokriging, GE-Kriging and GE-Cokriging. $*$ denotes gradient in $x$ direction and $**$ denotes gradient in $y$ direction. $\nabla$ denotes prediction of the gradient of QoI.}
    \label{tab:time_table}
\end{table}


\section{Conclusion}
\label{sec:concl}
In this work, we present a comprehensive gradient-enhanced multi-fidelity Cokriging method, namely GE-Cokriging,
which incorporates available gradient information of multi-fidelity data, i.e., low-fidelity and high-fidelity observation of QoIs and its gradients. We present several numerical examples to study the performance of GE-Cokriging. 
Our results show that GE-Cokriging can accurately predict the QoI and its gradients simultaneously.
We compare the performance of GE-Cokriging against GE-Kriging and multi-fidelity Cokriging, two popular GP-based prediction methods, and illustrate that GE-Cokriging is the most accurate, robust and efficient among these methods.

In particular, our result suggests that GE-Cokriging achieves better accuracy than GE-Kriging, this is because it exploits the information of the low-fidelity model. Also, GE-Cokriging yields more accurate results than using Cokriging for QoI and its gradients separately, because it takes advantage of the relation between these two quantities and makes use of corresponding data jointly.
Even when some of the low-fidelity gradient information is misleading, for example, the gradient of low-fidelity data is negative while that of high-fidelity data is positive, the GE-Cokriging method may still be robust enough to predict accurately on target functions with less uncertainty compared to those by Cokriging and GE-Kriging. Moreover, the GE-Cokriging helps to alleviate the singularity issue of the covariance matrix, which is quite common in GPR methods.
In terms of computational cost, the training of GE-Cokriging model, i.e., identifying hyperparameters, could take longer time than Cokriging in solving a high-dimensional problem, given that the dimension of the covariance matrix is expanded due to the incorporation of gradient samples. However, once these hyperparameters are specified, the QoI and its gradients can be predicted simultaneously. This saves total computational time compared with Cokriging, which requires constructing models for QoI and its gradients separately, and hence needs training at least two models.
Therefore, the overhead of training a model with a larger covariance matrix in GE-Cokriging is mitigated, and the overall time required to predict both QoI and its gradients for these three methods are comparable.

We note that our gradient-enhanced framework is also flexible for further extensions. In all of the numerical examples, we apply the commonly used stationary radial-basis function kernel. Other kernel functions, e.g., Mat\'ern kernels with different 
smoothness, can be used to solve problems with desired regularity constraints. In addition, non-stationary kernels can be applied in this framework to model heterogeneous systems more accurately. Another extension can be to relax the constraints on the sample data to address the situation of missing data. More specifically, in the numerical examples presented, the gradient information is available
with QoI at each observation location. Whereas in practice, it is possible that at some observation locations, either the QoI or its gradient is unavailable. In this scenario, modifications to the mean and covariance functions of the GP in our framework are needed. Moreover, we used the linear auto-regression form of the multi-fidelity Cokriging from~\cite{kennedy2000predicting}, which can be replaced by more general nonlinear auto-regression forms, e.g., the methods used in~\cite{perdikaris2017nonlinear, girolami2007data, lee2019linking}, or even the deep neural network, e.g.,~\cite{meng2020composite}. Finally, as we point out in Section~\ref{subsec:integral}, our framework can also be built based on the ``integral-enhanced'' perspective, which can be useful in specific practical problems.

\section*{Acknowledgments}
Yixiang Deng was supported by National Science Foundation (NSF) Award No. 1736088.
Xiu Yang was supported by the U.S. Department of Energy 
(DOE), Office of Science, Office of Advanced Scientific Computing Research
(ASCR) as part of Multifaceted Mathematics for Rare, Extreme Events in Complex
Energy and Environment Systems (MACSER).
Guang Lin gratefully acknowledges the support from National Science Foundation (DMS-1555072, DMS-1736364, and CMMI-1634832) and Brookhaven National Laboratory Subcontract 382247.


\bibliographystyle{plain} 
\bibliography{ref}

\end{document}